\documentclass[smallextended]{article}
\usepackage[utf8]{inputenc}
\usepackage{mathpazo}
\usepackage{relsize}
\usepackage{amsmath}
\usepackage{url}
\usepackage{physics}
\usepackage[round,numbers]{natbib}
\usepackage[margin=1in]{geometry}
\usepackage{amssymb}
\usepackage{standalone}
\usepackage{graphicx}
\usepackage{epstopdf}
\usepackage{hyperref}
\usepackage{subfigure}
\usepackage{cleveref}

\usepackage[font=small,labelfont=bf]{caption}
\usepackage{csquotes}
\usepackage{overpic}

\usepackage[normalem]{ulem}
\usepackage[rgb,dvipsnames]{xcolor}
\usepackage{bm}
\usepackage{booktabs}

\usepackage{multicol} 
\usepackage[math]{cellspace}
\AtBeginDocument{
    \heavyrulewidth=.08em
    \lightrulewidth=.05em
    \cmidrulewidth=.03em
    \belowrulesep=.65ex
    \belowbottomsep=0pt
    \aboverulesep=.4ex
    \abovetopsep=0pt
    \cmidrulesep=\doublerulesep
    \cmidrulekern=.5em
    \defaultaddspace=.5em
}
\definecolor{table1}{rgb}{0.9, 0.9, 0.98}
\definecolor{table2}{rgb}{0.8, 0.8, 1.0}

\newcommand{\dimx}{x^{*}}
\newcommand{\dimy}{y^{*}}
\newcommand{\dimR}{R^{*}}
\newcommand{\dimL}{L^{*}}
\newcommand{\dimwidth}{a^{*}}
\newcommand{\dimsep}{r^{*}}
\newcommand{\dimperm}{P_{\text{eff}}^{*}}

\newcommand{\eps}{\varepsilon}
\newcommand{\vecx}{\boldsymbol{x}}
\newcommand{\dimvecx}{\boldsymbol{x}^{*}}
\newcommand{\dimfullwidth}{w^{*}}
\newcommand{\dimfulll}{l^{*}}
\newcommand{\dimdif}{D^{*}}
\newcommand{\x}{x}
\newcommand{\y}{y}
\newcommand{\chanL}{L}

\newcommand{\dimt}{t^{*}}
\newcommand{\ndtime}{t}
\newcommand{\dimc}{c^{*}}
\newcommand{\charc}{c^{*}_0}
\newcommand{\conc}{c}
\newcommand{\dimdifspace}{\Omega^{*}}

\newcommand{\dimmem}{\Gamma^{*}}

\newcommand{\kin}{\kappa_{1}}
\newcommand{\kout}{\kappa_{2}}

\newcommand{\omone}{\omega_-}

\newcommand{\xb}{X}
\newcommand{\ybl}{Y_{1}}
\newcommand{\ybu}{Y_{2}}
\newcommand{\delfunc}{\hat{\delta}}
\newcommand{\zbl}{Z_{1}}
\newcommand{\zbu}{Z_{2}}
\newcommand{\inonex}{\bar{X}}
\newcommand{\inoney}{Y_-}
\newcommand{\inonez}{Z_-}
\newcommand{\inonezet}{\zeta_-}
\newcommand{\inthreey}{Y_+}
\newcommand{\inthreez}{Z_+}
\newcommand{\inthreezet}{\zeta_+}

\newcommand{\ybmatch}{Y_{i}}
\newcommand{\kmatch}{\kappa_{i}}

\newcommand{\amatch}{A_{i}}
\newcommand{\ina}{a_-}
\newcommand{\atild}{a_+}
\newcommand{\inb}{b_-}
\newcommand{\btild}{b_+}

\newcommand{\zinmatch}{Z_{\mp}}

\newcommand{\inymatch}{Y_{\mp}}
\newcommand{\timey}{\bar{y}}

\newcommand{\concbelow}{\conc_{-}}
\newcommand{\concabove}{\conc_{+}}

\begin{document}

\begin{center}
\rule{\textwidth}{0.4pt}\\[1em]
    {\Large \textbf{Effective permeability conditions for diffusive transport through impermeable membranes with gaps}}
\rule{\textwidth}{0.4pt}\\[1em]
    {\large Molly Brennan$^1$ \hspace{0.1cm} \textbullet \hspace{0.1cm} Edwina F. Yeo$^1$ \hspace{0.1cm} \textbullet  \hspace{0.1cm} Philip Pearce$^1$ \hspace{0.1cm} \textbullet \hspace{0.1cm} Mohit P. Dalwadi$^{1,2}$* } \\[1em]
    {$^{1}$Department of Mathematics, University College London, London, WC1E 6BT, UK}\\[1em]
    {$^{2}$Mathematical Institute, University of Oxford, Oxford, OX2 6GG, UK}\\[1em]
    {\small \texttt{*mohit.dalwadi@maths.ox.ac.uk}} \\[2em]
\end{center}

\begin{abstract}
Membranes regulate transport in a wide variety of industrial and biological applications. The microscale geometry of the membrane can significantly affect overall transport through the membrane, but the precise nature of this multiscale coupling is not well characterised in general. Motivated by the application of transport across a bacterial membrane, in this paper we use formal multiscale analysis to derive explicit effective coupling conditions for macroscale transport across a two-dimensional impermeable membrane with periodically spaced gaps, and validate these with numerical simulations. We derive analytic expressions for effective macroscale quantities associated with the membrane, such as the permeability, in terms of the microscale geometry. Our results generalise the classic constitutive membrane coupling conditions to a wider range of membrane geometries and time-varying scenarios. Specifically, we demonstrate that if the exterior concentration varies in time, for membranes with long channels, the transport gains a memory property where the coupling conditions depend on the system history. By applying our effective conditions in the context of small molecule transport through gaps in bacterial membranes called porins, we predict that bacterial membrane permeability is primarily dominated by the thickness of the membrane. Furthermore, we predict how alterations to membrane microstructure, for example via changes to porin expression, might affect overall transport, including when external concentrations vary in time. These results will apply to a broad range of physical applications with similar membrane structures, from medical and industrial filtration to carbon capture.
\end{abstract}

\section{Introduction}
Membrane transport plays an important role in many biological and industrial settings, including haemodialysis for virus filtration in laboratories \cite{haney2013separation}, kidney dialysis to treat renal failure \cite{cancilla_mathematical_2025}, carbon capture from flue gas \cite{aninwede_modeling_2025}, wastewater treatment \cite{dzygiel_chapter_2010} and bacterial communication \cite{tieleman_computer_2006}. In many of these applications transport across membrane structures, such as channels in the membrane, occurs over much smaller lengthscales than the overall system scales.  Membrane-mediated transport is therefore inherently multiscale, creating difficulties in resolving disparate lengthscales in mathematical modelling. A common way to deal with these difficulties is to use constitutive macroscale relationships motivated by physical intuition. These relationships typically link the flux through the membrane to the concentration difference across it, with standard conditions of the form
 \begin{equation}
   \frac{\partial\conc}{\partial n}\rvert_{\text{above}}= \frac{\partial \conc}{\partial n} \rvert_{\text{below}}=P_{\text{eff }}\left(\conc_{\text{above}} - \conc_{\text{below}} \right),\label{permeationeq}
\end{equation} 
where $c$ is the solute concentration, $\partial/\partial n$ is the derivative normal to the membrane, and $P_{\text{eff }}$ is a (typically fitted) constant of proportionality \cite{wong_cells_2009,melke_cell-based_2010,prajapati_how_2021,linares_silico_2015,schiesser_partial_2012}. Other authors have also incorporated the effect of membrane transport by coupling production and decay of concentrations on either side of the membrane \cite{barbarossa_mathematical_2016,goryachev_systems_2006}. While these constitutive conditions are generally successful in replicating observed phenomena in the parameter regimes in which their coefficients are estimated, and deal with the issue of computational complexity induced by disparate lengthscales, they may not be able to predict transport in new regimes because the lumped parameters within constitutive relations do not explicitly depend on microscale geometry.

To address the multiscale challenge of accounting for microscale channel effects in larger scale models, many authors have sought to derive effective conditions representing transport across membranes. A review of early work in this field is presented in \cite{wakeham_diffusion_1979}. Several authors consider homogenisation of a surface of periodically varying boundary conditions in order to describe effective membrane behaviour for various transport mechanisms, for example using (periodic) patches of perfect absorbers on an otherwise reflecting surface \cite{muratov_boundary_2008,bruna_effective_2015,belyaev_effective_1999,lindsay_first_2017,bernoff_boundary_2018}. With this method, one can derive effective boundary conditions for the bulk concentration in the region above, providing an effective uptake equation on one side of a membrane surface.

Transport across thin membranes has been studied in \cite{zampogna_effective_2020,zampogna_transport_2022}, where the authors consider advective transport (including diffusive transport and surface reactions in the latter paper), deriving appropriate effective jump conditions on the stress and solute flux across a membrane. In the absence of any surface reactions, the authors calculate that the leading-order concentration and concentration flux are continuous across the effective membrane surface to leading order. This is in part due to the thin membrane geometry that the authors investigate, which cannot sustain a leading-order concentration jump across the membrane. Here we are interested in a different class of membranes; those across which significant concentration differences can be sustained by membrane geometry alone. As such, we must include both a physical membrane with a significant thickness as part of our model as well as domains both above and below the membrane, in contrast to the thin membrane or single domain problems discussed above. Several authors have considered similar problems of membrane transport from a rigorous standpoint, proving convergence of specific effective problems and solution forms \cite{raveendran_upscaling_2022,marusic_asymptotic_2008,bhattacharya_effective_2022,del_vecchio_thick_1987}. These typically take the form of a derived microscale system to be solved, coupled to a macroscale problem, and provide mathematical insight into the structure of membrane transport problems. However, owing to their generality one typically cannot solve explicitly the microscale systems derived therein. Here, in order to gain further insight into the coupling between microscale geometry and macroscale function, we investigate a specific microscale geometry in which we can derive explicit relationships between scales.

We are primarily motivated by applications involving bacterial membranes -- more specifically, the outer membrane of gram-negative bacteria -- where transport is typically limited to specific channels with binding sites or non-specific channels, called porins, which provide a size-restricted passageway for small water-soluble molecules \cite{prajapati_how_2021}. Understanding transport across bacterial membranes is vital because the ability of bacterial populations to regulate complex collective behaviours that facilitate infection and resistance relies on the exchange and build up of small diffusible signal molecules that cross bacterial membranes \cite{miller_quorum_2001, dalwadi_emergent_2021}. The disruption of communication pathways is therefore a proposed target for interventions in preventing bacterial infection, particularly due to the growing numbers of antibiotic resistant bacteria \cite{naga2023time}. Transport is also fundamental in current prevention methods of bacterial infection, such as in antibiotic delivery, in which porins play a key role \cite{vergalli2020porins}. For many species of bacteria, antibiotics must cross bacterial membranes in order to be effective. In particular, transport across the outer membrane of gram-negative bacteria is complicated by many factors. For example, antibiotic molecules are often of a similar size to the porins and channels in the membrane so steric effects, including interaction with channel surfaces, can be a limiting factor in antibiotic permeation \cite{vergalli2020porins}. In addition, interactions between antibiotic molecules and binding sites in specific channels and the charge of the molecules can enhance transport \cite{prajapati_how_2021}. However, for transport of both signal and antibiotic molecules in bacterial populations, processes that occur on the scale of membrane channels can generate effects that occur over the colony scale, again highlighting the need for multiscale insight. Here we wish to specifically understand how membrane microstructure itself affects bacterial scale transport, independent of channel interactions and finite-size particle effects, focusing on diffusion of small solutes (typically with a molecular mass cut-off of around 600 Da) through general porins facilitating non-specific diffusion \cite{vergalli2020porins}.

In this paper, we use homogenisation via the method of multiple scales to derive explicit coupling conditions across an impermeable perforated membrane, allowing us to connect the concentration fields away from the membrane. Our membrane consists of an impermeable barrier periodically punctured with channels that are thin compared to their spacing. We consider both the limits in which gap height is much larger than gap width, which we call the long thin channel limit, and the limit in which gap height is around the same size as gap width, which we refer to as the $\mathcal{O}(1)$ aspect ratio limit. The multiscale (asymptotic) techniques we use to derive the effective coupling conditions to apply across our membrane are similar to those used to derive effective coupling conditions in wave transmission problems in Faraday cages in \cite{chapman_mathematics_2015,hewett2016homogenized}. More generally, boundary homogenisation methods are widely used to obtain coupling conditions for wave transmission problems across thin interfaces \cite{delourme2012approximate,delourme2013well,marigo2016homogenization}.  We exploit the extreme disparity in the spatial scales in our problem, with our derived effective coupling conditions formally incorporating channel lengthscale effects over bacterial lengthscales. Our results allow one to formally treat the membrane as a continuous boundary with effective conditions which couple the concentration on either side, and allow us to understand how these effective conditions depend on the microscale properties of the membrane. We derive coupling conditions for both steady and unsteady transport across the membrane. Notably, the coupling conditions we derive for the steady problem are mathematically equivalent to the standard permeability equation \eqref{permeationeq}, and we are able to explicitly determine the effective permeability $P_{\text{eff}}$ in terms of the microscale membrane geometry. In the unsteady case, we derive new effective coupling conditions, which gain terms with integrals in time i.e. generating emergent memory properties.
We use our results to investigate how the microscale structure of the membrane impacts its effective permeability. Our analysis allows us to understand in which limits the constitutive models are valid, when one might expect them to break down, and the appropriate form of the coupling conditions in these scenarios.

The structure of this paper is as follows. In \S \ref{s2mathframe} we introduce the mathematical description of our problem, and discuss the asymptotic structure and specific asymptotic regions of the problem. In \S \ref{longthin} we consider the steady problem, and, using boundary layer theory and homogenisation via the method of multiple scales, we systematically derive the appropriate effective coupling conditions. For clarity, we present analysis of the long thin channel limit in the main text, and present analysis of the $\mathcal{O}(1)$ aspect ratio limit in Appendix \ref{o1details}. In \S \ref{s4timedep} we consider the unsteady problem, allowing us to determine when it is appropriate to use the steady effective conditions we derive in \S \ref{longthin}, and the appropriate unsteady conditions to use when it is not. We also present an early-time approximation to our effective coupling conditions, when they are generically slow to converge, using an asymptotic Euler-Maclaurin summation. In \S \ref{numerics} we validate our effective results against simulations of the full membrane geometry. In \S \ref{permregimes} we explore the implications of  our results and compare them to other models for membrane and channel transport. Finally, in \S \ref{discuss} we discuss our results, and highlight how they could be used in other applications.

\section{Mathematical framework} \label{s2mathframe}
\begin{figure}[t]
    \centering
    \includegraphics[width=\textwidth]{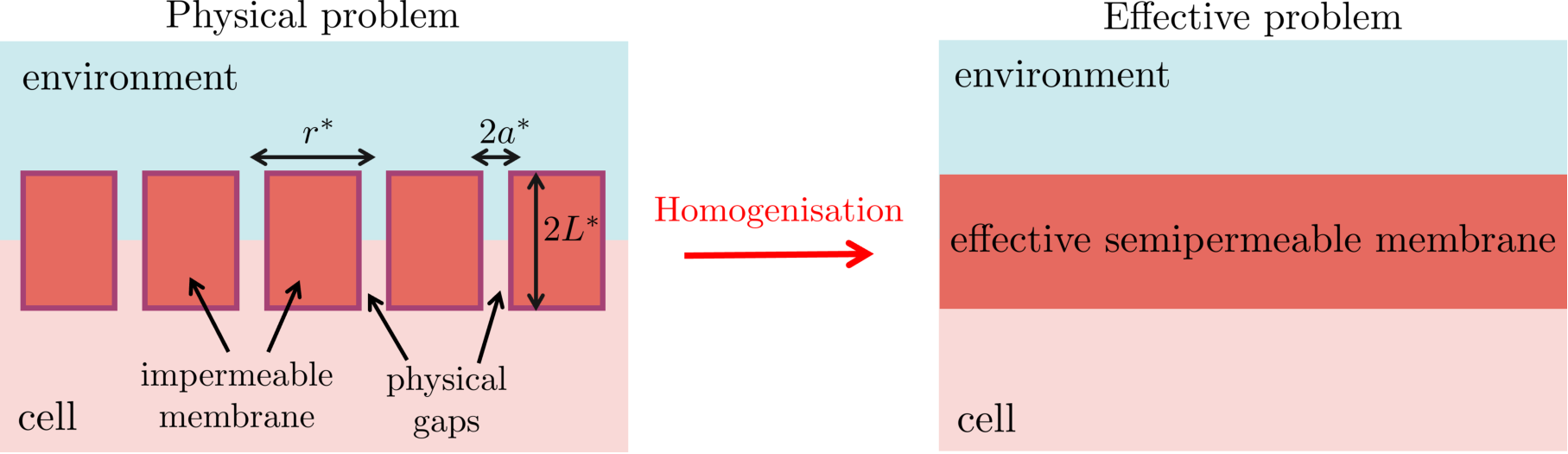}
    \caption{Schematic demonstrating the output of our homogenisation procedure. We systematically derive effective coupling conditions that allow us to accurately model the membrane as a continuous and homogeneous boundary, removing the need to account for each gap individually. Dimensional parameters are defined in \S \ref{s2mathframe}.}
    \label{fig:overview}
\end{figure}

We consider the diffusive transport of a solute, representing e.g.~autoinducers or antibiotics, in the 2D domain $\dimdifspace = \{ (\dimx, \dimy) \in \mathbb{R}^2 \}$ with the membrane modelled as an impermeable wall with finite thickness, $2\dimL$, periodically perforated with physical gaps of width $2\dimwidth$ each separated by a distance of $\dimsep$. These gaps model non-specific channels, i.e.~channels with no interaction or binding, across which a solute consisting of small molecules diffuses freely. Mathematically we describe the impermeable part of the membrane (i.e. the membrane excluding the gaps) as $\dimmem = \{\dimvecx=(\dimx, \dimy): \dimx \in \mathbb{R}, \dimy \in (-\dimL,\dimL) \} \backslash \{ \dimvecx=(\dimx, \dimy): \dimx \in (\dimsep j - \dimwidth, \dimsep j + \dimwidth), \dimy \in (-\dimL, \dimL), j \in \mathbb{Z} \}$
with $\partial \dimmem$ denoting its boundary. We denote our solute concentration by $\dimc(\dimvecx,\dimt)$, where $\dimvecx$ is the spatial vector coordinate and $\dimt$ is time. We will use homogenisation via the method of multiple scales to derive effective coupling conditions across the membrane (illustrated schematically in Figure \ref{fig:overview}). The problem we consider can be thought of as the `inner' problem (in the language of matched asymptotic expansions) in which we derive the correct coupling conditions for an unsteady `outer' problem which sees the membrane as a curved interface. As such, we consider a locally flat membrane for the inner problem, noting that our results will be valid for membranes with moderate curvature in the outer problem.

We consider diffusive transport through the impermeable membrane. Mathematically this is given by
\begin{subequations}
\label{dimensionalsystem}
\begin{align}
    &\frac{\partial \dimc}{\partial \dimt} = \dimdif \nabla^{*2} \dimc,& \hspace{1cm} &\dimvecx \in \dimdifspace\backslash \dimmem,& \label{dimfull1}\\
    &\frac{\partial \dimc}{\partial n} = 0,&  &\dimvecx \in \partial \dimmem,\label{dimfull2}\\\
    &\dimc(\dimvecx,0) = \dimc_{\text{init}}(\dimvecx),& &\dimvecx \in \dimdifspace \backslash \dimmem,& \label{dimfull3}
    \end{align}
\end{subequations}
where $\dimc_{\text{init}}$ denotes a general initial condition, $\dimdif$ is the constant diffusivity of the solute and ${\partial}/{\partial n}$ denotes the derivative in the direction of the outward pointing normal to the membrane. As we assume no solute is absorbed on the impermeable membrane walls, \eqref{dimfull2} describes no flux boundary conditions. While far-field conditions ($\dimy \to \pm \infty$) are also required to fully close the problem, our analysis is agnostic to their specific form for non-pathological cases and so we do not prescribe these for generality. Our overarching goal is to derive the effective coupling conditions that link the \enquote{internal} concentration as $\dimy \rightarrow - \infty$, below the membrane, to the \enquote{external} concentration at $\dimy \rightarrow \infty$ above the membrane.

We provide typical parameter values for bacterial systems, which motivate our specific geometry, in Table \ref{tab:par}. 
 \begin{table}[t!]
 \centering
 \begin{tabular}{|l|l|l|l|}
 \hline
 \textbf{Dimensional parameter} & \textbf{Physical interpretation} & \textbf{Size (nm)} & \textbf{Reference}\\
 \hline
 $\dimR$ & macroscale length (e.g. cell radius) & $360-3000$ & \cite{bakshi_superresolution_2012,hau_ecology_2007} \\
 \hline
 $\dimwidth$ & porin radius (half porin width) & $0.325 - 4$ & \cite{prajapati_how_2021,benn_phase_2021}\\
 \hline
 $2\dimL$ & porin length (membrane thickness) & $2.4-12$ & \cite{niederweis_mycobacterial_2003,mai-prochnow_gram_2016,bayer_zones_1991} \\
 \hline
 $\dimsep$ & centre-to-centre porin spacing &$9-10.8$ &\cite{amako_electron_1996,benn_phase_2021}\\
 \hline
 \end{tabular}
 \caption{Dimensional membrane parameters for a range of bacterial species.}
 \label{tab:par}
 \end{table}
In general, bacteria can regulate the expression of porins, for example, lowering expression in response to environmental factors to reduce permeability to e.g. antibiotics. They can also alter porin configuration, changing them to a closed configuration in response to voltage, acidic pH and binding of certain molecules \cite{delcour_outer_2009,benn_phase_2021}. However, for simplicity, here we consider purely static membrane geometries neglecting temporal effects in porin expression.   

We nondimensionalise the problem using the following macroscale scalings 
\begin{equation}
    \dimt = \frac{{\dimR}^2}{D} \ndtime, \hspace{1cm} \dimvecx = \dimR \vecx, \hspace{1cm} \dimc,  = \charc\conc, \hspace{1cm} \dimc_{\text{init}} = \charc \conc_{\text{init}}, \label{nondiming}
\end{equation}
where $\dimR$ is a characteristic macroscale length (e.g. cell radius) and $\charc$ is a characteristic macroscale concentration. 
Applying our nondimensionalisation, \eqref{nondiming}, to our governing equations, \eqref{dimensionalsystem}, we have
\begin{subequations}
\label{nondimgoverning}
\begin{align}
    &\frac{\partial \conc}{\partial \ndtime} = \nabla^2 \conc& \hspace{1cm} &\vecx \in \Omega \backslash \Gamma, \label{multfull1}\\
    & \frac{\partial \conc}{\partial n} = 0& &\vecx \in \partial \Gamma,\label{multfull2}\\
    &\conc(\vecx, 0) = \conc_{\text{init}}(\vecx)& &\vecx \in \Omega \backslash \Gamma. \label{multfull3}
\end{align} 
\end{subequations}
Our dimensionless membrane is now defined as
\begin{equation}
    \Gamma = \{\vecx = (x, y): x \in \mathbb{R}, y \in (- L, L) \} \backslash \{\vecx = (x,y): x \in (\delta j - \delta \eps, \delta j + \delta \eps), y \in (-\chanL, \chanL), j \in \mathbb{Z}\},
\end{equation}
where we have introduced the rescaled parameters
\begin{equation}
   \delta = \frac{\dimsep}{\dimR}, \hspace{1cm}  \eps= \frac{\dimwidth}{\dimsep}, \hspace{1cm}  L = \frac{\dimL}{\dimR}. \label{nondimpars}
\end{equation}
These parameters $\delta$, $\delta \eps$ and $\chanL$ now represent the nondimensional channel spacing, channel width and membrane thickness respectively. We illustrate our full nondimensional geometry with the membrane and porins alongside the asymptotic structure of our problem in Figure \ref{fig:regionsmultipore}.

To get an idea of the relative sizes of our nondimensional parameters we take, for example, $\dimR=360\;$nm, $\dimsep=10\;$nm, $\dimwidth=0.325\;$nm and $2\dimL = 12\;$nm, from Table \ref{tab:par} giving approximate scalings for our nondimensional parameters of $\delta = 0.028$, $\eps = 0.0325$ (i.e. $\delta \eps = 0.00091$)  and $\chanL = 0.017$. With this choice of physical parameter values, we have $\delta \eps \ll \chanL$ and $\delta, \eps \ll 1$, meaning adjacent channels are close relative to bacterial radius, channel width is small compared to channel spacing and channels are much longer than they are wide, motivating the asymptotic structure we detail below. 

\subsection{Asymptotic structure}
We proceed in deriving an effective coupling condition for transport across the membrane by exploiting the disparity in lengthscales within the problem. Specifically, we consider the scenario in which the membrane has many well-separated gaps, i.e gap width is much smaller than the separation length. Mathematically the limit of many gaps corresponds to $\delta \ll 1$, and the limit of well separated gaps corresponds to $\eps \ll 1$. From Table \ref{tab:par} we see that ${L}/{\delta \eps} \approx 1-100$. The limit in which ${L}/{\delta \eps} = \mathcal{O}(1)$, which we refer to as the $\mathcal{O}(1)$ aspect ratio limit, is of potential physical relevance, and acts as a distinguished limit here, with the \enquote{long thin channel} limit, $\delta \eps \ll L$, emerging as a sublimit of this problem. However, we later demonstrate that the coupling conditions derived in the long thin channel problem work well across most parameters investigated, since this captures transport  dominated by channel entrance effects as well as channel length effects. As such, we present the long thin channel limit in the main text, allowing for more straightforward mathematical analysis and more interpretable derived effective coupling conditions. We present the $\mathcal{O}(1)$ aspect ratio limit in Appendix \ref{o1details}, and discuss the key results for this limit in \S \ref{o1limit}. 

\begin{figure}[t]
\centering
    \includegraphics[width =0.8\textwidth]{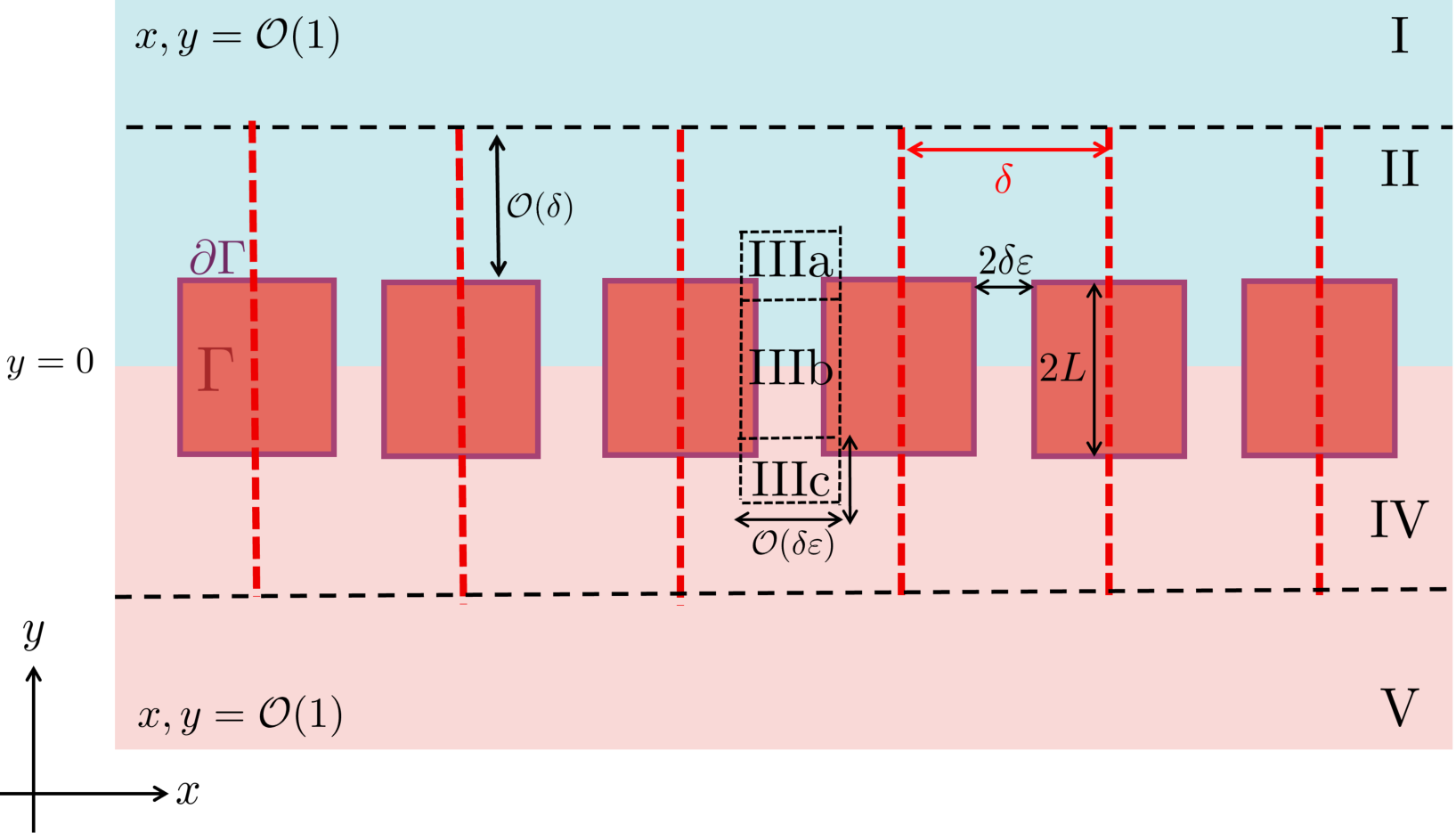}
    \captionof{figure}{Full asymptotic structure and nondimensional geometry of the perforated membrane problem in 2D. We show $\Gamma$, the impermeable parts of membrane in red, and its boundary, $\partial \Gamma$ in purple, where we apply no flux conditions. Regions \mbox{I} and \mbox{V} denote our outer regions where $x$ and $y$ are $\mathcal{O}(1)$. Regions \mbox{II} and \mbox{IV} denote the boundary layers on either side of the membrane and regions \mbox{III}a, b and c represent our three inner regions. In the $\mathcal{O}(1)$ aspect ratio problem these collapse to one inner region.}
    \label{fig:regionsmultipore}
    \end{figure}

\begin{figure}[t]
\centering
\begin{overpic}[abs,unit=1mm,scale=0.45,grid=false]{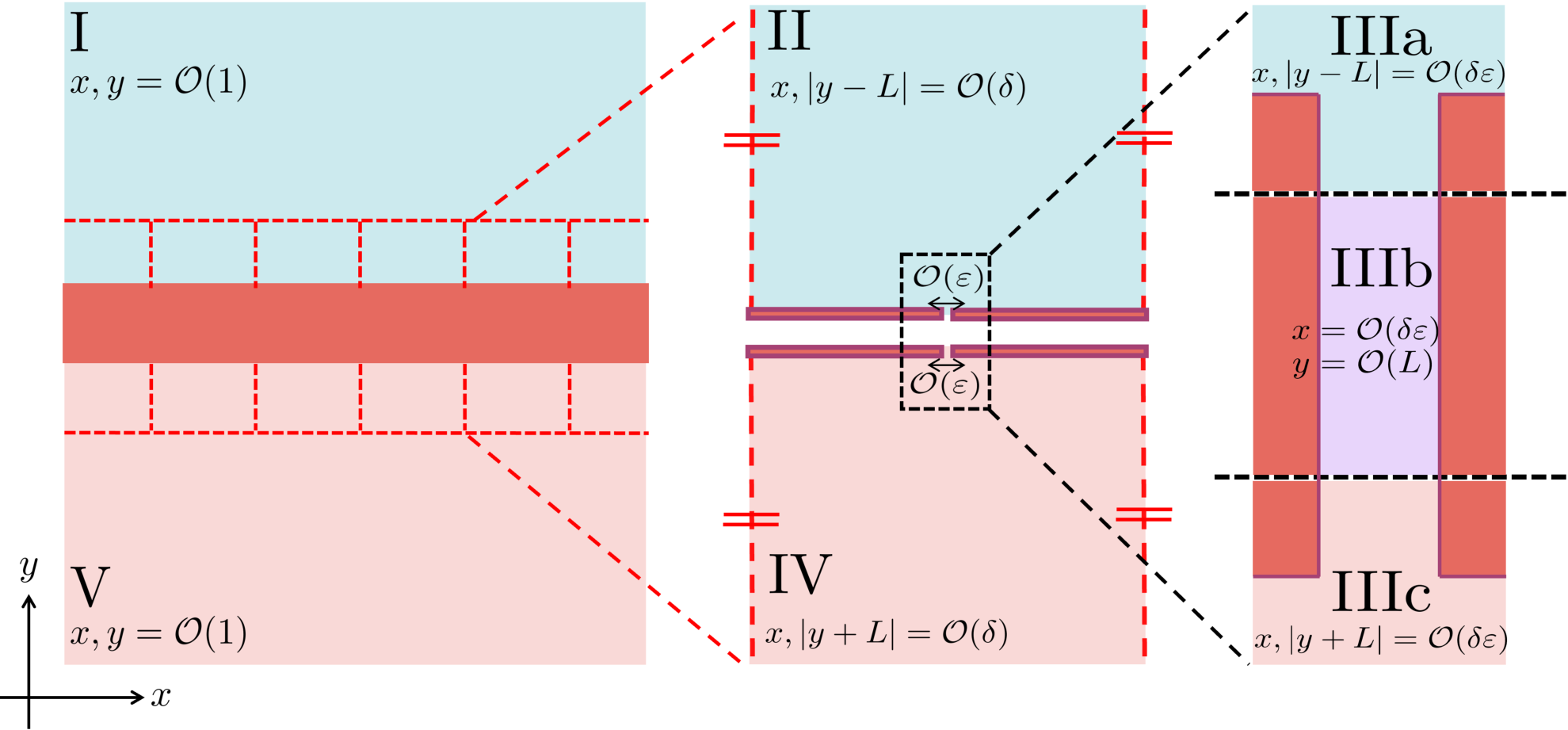}
\put(0,75){(a)}
\put(65,75){(b)}
\put(115,75){(c)}
\end{overpic}
\captionof{figure}{Asymptotic structure of (a) the outer regions, (b) the boundary layer region cell problem and (c) the inner regions of the multiple gaps problem in 2D. In each region we show the asymptotic size of the $(x,y)$ coordinates.}
\label{fig:asymptoticstructures}
\end{figure}

In both of the limits described above, our domain has several distinct asymptotic regions in space, as highlighted in Figure \ref{fig:regionsmultipore} and detailed in Figure \ref{fig:asymptoticstructures}. In the order we discuss them below, we term these the `outer', `boundary', and `inner' regions. In the outer regions away from the membrane, $\vecx = \mathcal{O}(1)$, and these are the regions we wish to connect via our effective coupling conditions. Our outer regions \mbox{I} and \mbox{V} are shown in Figure \ref{fig:asymptoticstructures} (a). We then have boundary layer regions, \mbox{II} and \mbox{IV}, on either side of the membrane where $\y = \mathcal{O}(\delta)$. In these regions the individual nature of the gaps becomes evident, but we still cannot see their full structure and the gaps act as effective sinks or sources. In these regions, we must analyse a multiple scales problem in $\x$ with long-scale variation of $\mathcal{O}(1)$ and short-scale variation of $\mathcal{O}(\delta)$, giving us periodic mathematical `cells' of $\mathcal{O}(\delta)$ as shown in Figure \ref{fig:asymptoticstructures}(b). Our two boundary layer regions are connected by our inner regions labelled \mbox{III}a-c where $x = \mathcal{O}(\delta \eps)$, shown in \ref{fig:asymptoticstructures}(c). Here we can see the full finite geometry of the gap. Inner regions \mbox{III}a and \mbox{III}c describe the $\mathcal{O}(\delta \eps)$ regions about our two gap openings, and inner region \mbox{III}b describes the length of the channel where $y = \mathcal{O}(\chanL)$. In the $\mathcal{O}(1)$ aspect ratio limit \mbox{III}a-c coalesce into a single inner region.

\section{Steady problem} \label{longthin}
We start by considering the steady problem, setting the time derivative in \eqref{multfull1} to zero. To determine our effective coupling conditions, we calculate the leading-order concentration in each of our asymptotic regions and then asymptotically match our solutions across each region. This process connects the outer region concentrations, systematically taking into account the variation across the channel due to geometry. As discussed above we present the full solution for the long thin channel limit, $\delta \eps \ll \chanL$.

For convenience we first analyse the inner regions, then the boundary layer regions, and finally the outer regions (see Fig.~\ref{fig:asymptoticstructures}(c)). Our goal is to understand how the flux through these regions depends on the problem geometry and how this information generates an appropriate effective macroscale membrane condition.

\subsection{Inner regions} \label{longthininner}
We start by considering the three inner regions in our periodic cell, as shown in Figure \ref{fig:asymptoticstructures}(c). In these inner problems we have $\x = \mathcal{O}(\delta \eps)$. Starting within the channel and working outwards, we first consider region \mbox{III}b, in which concentration diffuses across the length of the channel.

\subsubsection{Region \mbox{III}b} \label{channellength}
The first and simplest region is Region \mbox{III}b where we have $\x = \mathcal{O}(\delta \eps)$ and $\y = \mathcal{O}(L)$. Although $\delta \eps \ll \chanL$ here, the precise size of $\chanL$ beyond this is not important for our analysis. As it is mathematically convenient to do so, we henceforth formally treat $L=\mathcal{O}(1)$, which does not affect our leading-order analysis. We scale into region \mbox{III}b by defining a rescaled horizontal coordinate
\begin{equation}
    \x = \delta \eps \inonex,
\end{equation}
which transforms \eqref{nondimgoverning} into
\begin{subequations}
\label{in2governing}
\begin{align}
    &\frac{\partial^2 \conc}{\partial \inonex^2} + (\delta \eps)^2 \frac{\partial^2 \conc}{\partial \y^2} =0,& \hspace{1cm} &|\inonex|<1, |\y|< \chanL, \label{in2eq1}\\
    &\frac{\partial \conc}{\partial \inonex} = 0,& & |\inonex|=1, |\y|< \chanL,\label{in2eq2}
\end{align}
\end{subequations}
where the asymptotic Region \mbox{III}b is defined via $\{(\inonex,y): |\inonex| < 1, |\y| < \chanL \} $. At leading-order, \eqref{in2governing} implies that $\conc \sim \conc(\y)$. We determine this leading-order function of $\y$ by integrating \eqref{in2eq1} over $\inonex \in (-1,1)$ and applying the boundary condition \eqref{in2eq2}. Hence, the leading-order concentration satisfies
\begin{align}
\dfrac{\mathrm{d}^2\conc}{\mathrm{d} y^2} = 0,
\end{align}
effectively corresponding to one-dimensional diffusion through the channel.
We may therefore deduce
\begin{equation}
    \conc \sim \alpha \y + \beta, \label{steadsolin2}
\end{equation}
where the integration constants $\alpha$ and $\beta$ must be determined by asymptotic matching.  We next consider Regions \mbox{III}a and \mbox{III}c, the gap opening region. We emphasise that these smaller asymptotic regions terminate Region \mbox{III}b at finite values of $y$, rather than transitioning at asymptotically large values of $y$.

\subsubsection{Regions \mbox{III}a and \mbox{III}c} \label{inner1and3}
We now investigate Region \mbox{III}c, which is defined by an $\mathcal{O}(\delta \eps)$ region about the gap opening centred at $\vecx = (0,-\chanL)$. The analysis for Region \mbox{III}a will follow from the Region \mbox{III}c analysis immediately from symmetry. We scale into Region \mbox{III}c defining rescaled horizontal and vertical coordinates
\begin{equation}
    \x = \delta \eps \inonex,  \hspace{1cm} \y = -\chanL + \delta \eps \inoney, \label{inreg1scal}
\end{equation}
converting \eqref{nondimgoverning} into the following scaled system
\begin{subequations}
\label{in1fulleq}
    \begin{align}
    &0=\frac{\partial^2 \conc}{\partial \inonex^2} + \frac{\partial^2 \conc}{\partial \inoney^2} ,& \hspace{1cm} &\{\inonex \in \mathbb{R}, \inoney <0\} \cup \{|\inonex|<1, \inoney>0\}, \label{in1eq1full}\\
    &\frac{\partial \conc}{\partial n} = 0,& &\{|\inonex|>1, \inoney=0\} \cup \{|\inonex|=1,  \inoney>0\}.\label{in1eq2full}   
    \end{align}
\end{subequations}
Although we are formally considering an asymptotic series expansion of our concentration, only the leading-order problem will be relevant for our purposes here so we work directly with \eqref{in1fulleq}. 

\begin{figure}[t]
\centering
    \includegraphics[width =\textwidth]{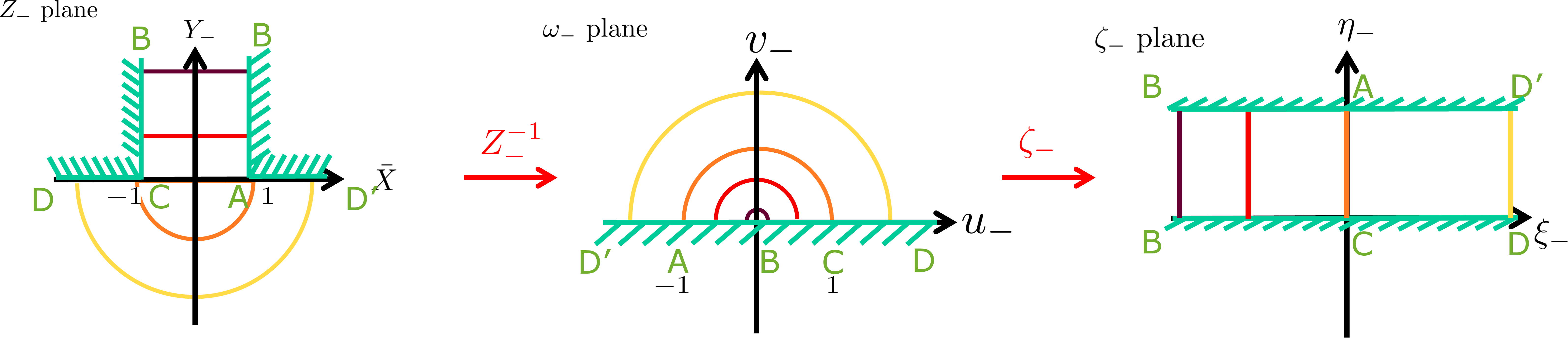}
    \captionof{figure}{Schematic to show the effects of the conformal maps $Z_{-}^{-1}$, \eqref{schwarzmap}, and $\zeta_{-}$, \eqref{zeta1map}, on the physical domain, with contour lines to show how the space is transformed.}
    \label{fig:shwarzchrist}
\end{figure}

As \eqref{in1eq1full} represents Laplace's equation in a 2D domain, we use conformal mapping to map our Region \mbox{III}c problem into a domain in which it is easier to solve. Specifically, setting $\inonez= \inonex + i \inoney$, we use the combination of  a Schwarz--Christoffel map from the upper half-plane to the open polygon representing our physical domain and a map from the upper half-plane to an infinite strip, in which our problem is straightforward to solve \cite{carrier_functions_2005}. These two maps are shown schematically in Figure \ref{fig:shwarzchrist}. Our Schwarz-Christoffel map from the upper half-plane to the physical domain takes the form
\begin{equation}
    \inonez = -1 - \frac{2}{\pi} \left( \left( \omone^2 - 1 \right)^{\frac{1}{2}} - \frac{1}{2 i} \log \left( \frac{i - \left( \omone^2 - 1 \right)^{\frac{1}{2}}}{i + \left( \omone^2 - 1 \right)^{\frac{1}{2}}} \right) \right), \label{schwarzmap}
\end{equation}
where we chose the branch cuts to lie outside our physical domain for the complex logarithm and square root functions. 

Combining \eqref{schwarzmap} with the standard map from the upper half-plane to the semi-infinite strip 
\begin{equation}
\inonezet = \frac{2}{\pi} \log{\omone}, \label{zeta1map}  
\end{equation}
where $\inonezet = \xi_{-} + i \eta_{-}$, we can write the full transformation from the semi-infinite strip to our physical domain as
\begin{equation}
 \inonez = -1 - \frac{2}{\pi} \left( \left( e^{\pi \inonezet} - 1 \right)^{\frac{1}{2}} - \frac{1}{2 i} \log \left( \frac{i - \left( e^{\pi \inonezet} - 1 \right)^{\frac{1}{2}}}{i + \left( e^{\pi \inonezet} - 1 \right)^{\frac{1}{2}}} \right) \right). \label{full map}   
\end{equation} 
In the semi-infinite strip, our problem is now given by
\begin{subequations}
\begin{align}
    &\dfrac{\partial^2 \conc}{\partial \eta_{-}^2} + \dfrac{\partial^2 \conc}{\partial \xi_{-}^2} = 0,& \hspace{1cm} &\{ \xi_{-} \in \mathbb{R}, 0 < \eta_{-} < 2 \}, \label{in1zeta1}\\
    &\frac{\partial \conc}{\partial \eta_{-}} = 0,& &\eta_{-} = 0,2. \label{in1zeta2}
\end{align}
\end{subequations}

Our remaining task is to find solutions to Laplace's equation with a finite flux through the channel region. In the transformed $\inonezet$ plane, the general solution corresponding to a finite flux is $\conc \sim \ina \Re{\inonezet} + \inb $, which can be simply transformed into the $\omone$ plane using \eqref{zeta1map}:
  \begin{equation}
     \conc \sim  \frac{2}{\pi} \ina \Re{\log{\omone}} + \inb, \label{steadsolin1}
 \end{equation}
where $\ina$ and $\inb$ are constants to be determined via later matching. 

For these matching purposes, it is helpful to record various far-field behaviours. For the form of the transformation \eqref{schwarzmap} we may deduce that  
\begin{equation}
\inonez \sim - \frac{2}{\pi} \omone \rightarrow - \infty \quad \text{as} \hspace{0.25cm} |\omone| \rightarrow \infty. \label{zlimominf}   
 \end{equation}
Therefore, combining \eqref{steadsolin1} and \eqref{zlimominf}, we have
\begin{equation}
    \conc \sim 
    \frac{2 \ina}{\pi}\log |\inonez| + \frac{2\ina}{\pi} \log{\frac{\pi}{2}} + \inb \quad \text{as} \hspace{0.25cm}  |\inonez| \rightarrow  \infty, Y_{-} < 0. \label{in3ctobl}
\end{equation}
Similarly, from \eqref{schwarzmap}, we deduce that
\begin{equation}
\inonez \sim - 1 -\frac{2 i}{\pi} - \frac{2 i}{\pi} \log{\omone} + \frac{2 i}{\pi} \log{2} \quad \text{as} \hspace{0.25cm}  \omone \rightarrow 0, \label{zlimom0} 
\end{equation}
and then combining \eqref{steadsolin1} and \eqref{zlimom0} we have
\begin{equation}
    \conc \sim - \ina \inoney - \frac{2 \ina}{\pi} + \frac{2 \ina}{\pi} \log{2} + \inb \quad \text{as} \hspace{0.25cm} \inonez \rightarrow i \infty. \label{in3ctoin3b}
\end{equation}

As noted above, the analysis for Region \mbox{III}a follows immediately via symmetry. Very briefly, we scale into this region using 
\begin{equation}
    \x = \delta \eps \inonex \hspace{1cm} \y = \chanL + \delta \eps \inthreey. \label{inreg3scal}
\end{equation}
Setting $\inthreez= \inonex + i \inthreey$, the equivalent map taking the semi-infinite strip in the $\inthreezet$ plane to the physical $\inthreez$-plane is
\begin{equation}
    \inthreez = 1 + \frac{2}{\pi} \left( \left( e^{\pi \inthreezet} - 1 \right)^{\frac{1}{2}} - \frac{1}{2 i} \log \left( \frac{i - \left(e^{\pi \inthreezet} - 1 \right)^{\frac{1}{2}}}{i + \left(e^{\pi \inthreezet} - 1 \right)^{\frac{1}{2}}} \right) \right). \label{mapporeupper}
\end{equation}
Hence, our solution in Region \mbox{III}a is
\begin{equation}
    \conc \sim \atild \Re{\inthreezet} + \btild \label{steadsolin3} 
\end{equation}
where $\atild$ and $\btild$ are determined by later asymptotic matching, and $\inthreezet $ is our intermediate variable describing the semi-infinite strip. In this case the far-field behaviour takes the forms
\begin{subequations}
\begin{align}
    \conc &\sim  \frac{2 a_{+}}{\pi} \log{|Z_{+}|} + \frac{2 a_{+}}{\pi}\log{\frac{\pi}{2}} + b_{+}, \quad \text{as } |Z_{+}| \rightarrow \infty, Y_{+} >0. \label{in3atobl}\\
\conc &\sim a_{+} Y_{+} - \frac{2 a_{+}}{\pi} + \frac{2 a_{+}}{\pi}\log{2}+ b_{+}, \quad \text{as } Z_{+} \rightarrow -i \infty. \label{in3atoin3b}
\end{align}
\end{subequations}

We have now determined the solutions in the three inner regions. Our remaining tasks are (i) to derive the leading-order concentration in the boundary layer regions, where the individual nature of the channels is apparent as point sinks or sources, and (ii) match between all the asymptotic regions to obtain the effective coupling conditions we seek.

\subsection{Boundary layers: Regions \mbox{II} and \mbox{IV}} \label{BLsec}
As with our inner regions, the analysis for Region \mbox{II} follows from the Region \mbox{IV} analysis via symmetry. Region \mbox{IV} arises when $\y = \mathcal{O}(\delta)$  with $\y<-L$ as shown in Figure \ref{fig:asymptoticstructures}(b). The local periodicity of this problem means that this takes the form of a multiple scales problem in $\x$ where we have periodic delta sinks or sources situated at $\delta j$ for $j \in \mathbb{Z}$ on an otherwise impermeable line. We choose our coordinates so that the centre of one gap lies on $\vecx = \vb{0}$, and scale into our periodic cell-level problem on this gap centred at the origin, using the scaling
\begin{equation}
  \y = -\chanL + \delta \ybl,  \label{bl1scale1}
\end{equation}
and introducing the microscale variable 
\begin{equation}
    \xb = \frac{\x}{\delta}.
\end{equation}
In the standard multiple scales manner, we assume that $\x$ and $\xb$ are independent of one another, considering $\conc = \conc(\x, \xb, \ybl)$. To remove the extra degree of freedom this introduces, we impose periodicity in $\xb$. The symmetry of the geometry means that this periodicity is equivalent to imposing no flux boundary conditions on the periodic boundaries of our cell.

To match the Region \mbox{III}a and \mbox{III}c solutions, \labelcref{steadsolin1,steadsolin3}, into our boundary layer Regions \mbox{II} and \mbox{IV}, we have to match the flux from the inner problems through regions of $\mathcal{O}(\eps)$ in the boundary layer problems. To facilitate this matching we introduce delta functions in the leading-order boundary layer problems, representing the inner regions as effective point sinks or sources. The strength of these effective delta functions will be related to the microscale channel geometry and flux through later matching. \footnote{A strictly formal asymptotic matching procedure would require writing $c$ as an asymptotic expansion and taking care about the asymptotic strength of the delta function. However, since we are able to solve the problem with an $\mathcal{O}(1)$-strength delta function, it is more straightforward to solve this general problem without considering relative strengths, and then to determine the precise strength of the delta function \emph{a posteriori} via matching.}

\begin{figure}[t]
\centering
    \includegraphics[width = 0.7\textwidth]{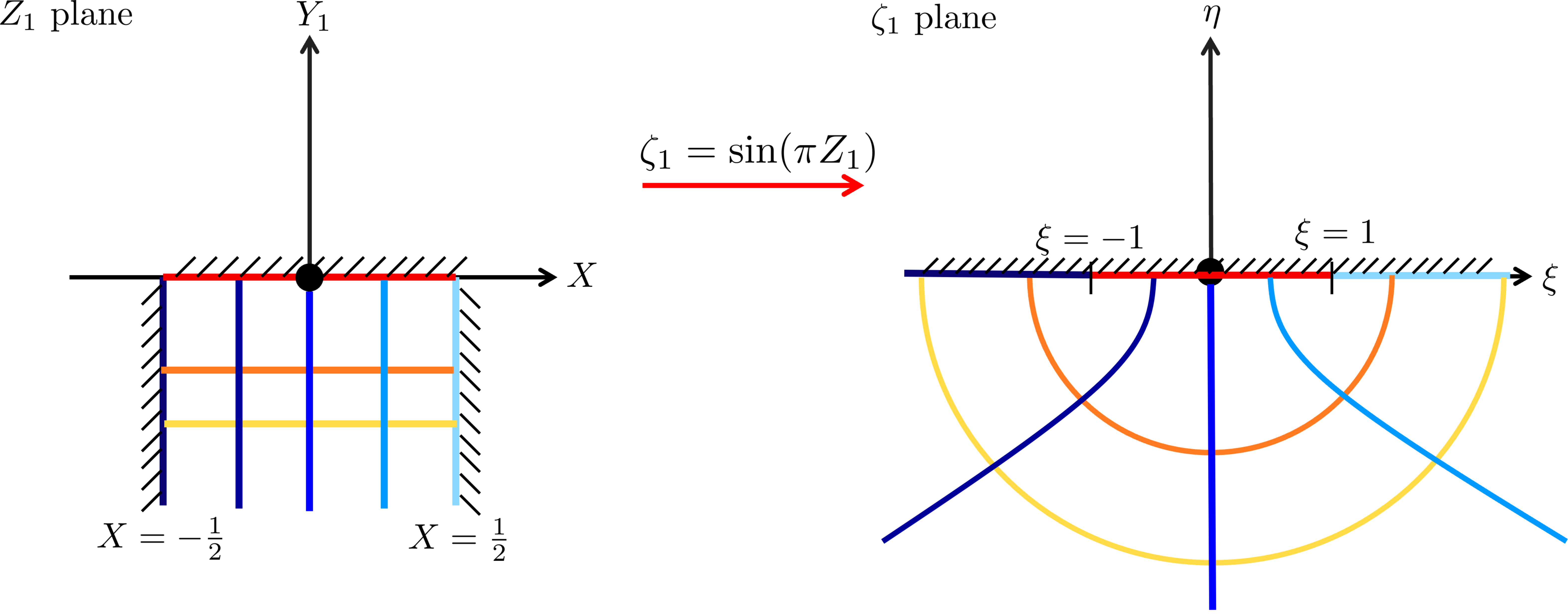}
    \captionof{figure}{Cell level Region \mbox{IV} under the mapping \eqref{confmapb2h} showing the transformation of lines of constant $X$ and $Y_1$.}
    \label{boundarymap1}
\end{figure}

The leading-order problem in Region \mbox{IV} is therefore

\begin{subequations} \label{leadb1}
\begin{align}
    &\nabla_{B1}^2 \conc = 0,& \hspace{0.5cm} &\ybl <0,\quad  -\frac{1}{2} < \xb < \frac{1}{2}, \label{leadb1e1}\\
    &\frac{\partial \conc}{\partial \ybl} = \kin \delfunc(\xb, \ybl) = \kin \delfunc(\xb) ,& & \ybl =0, \quad -\frac{1}{2} < \xb < \frac{1}{2},\label{leadb1e2} \\
     &\frac{\partial \conc}{\partial \xb} = 0,& &\ybl < 0, \quad \xb = \pm \frac{1}{2}, \label{leadb1e3} 
\end{align}
\end{subequations}
where we denote $\nabla_{B1}^2:= \frac{\partial^2 }{\partial \xb^2} + \frac{\partial^2 }{\partial \ybl^2}$ to represent the Laplacian purely in microscale variables. Here, $\delfunc$ is the Dirac delta function introduced for matching purposes as discussed above, and $\kin$ is the (as-of-yet unknown) strength of the effective source/sink given by the flux, determined later via matching. We have replaced the condition of periodicity in $\xb$ with a no flux condition, which is equivalent due to the symmetric geometry of this cell problem. Setting $\zbl = \xb + i \ybl$, we use the standard conformal map 
\begin{equation}
    \zeta_{1} = \sin(\pi \zbl), \label{confmapb2h}
\end{equation}
to map our open box domain to the lower half-plane, as shown in Figure \ref{boundarymap1}.

In the $\zeta_{1}$ plane, \eqref{leadb1} becomes
\begin{subequations} \label{zetaeqs}
\begin{align}
    &\nabla^2_{\zeta 1}\conc = 0,&  \hspace{1cm} &\eta < 0, \xi \in \mathbb{R},\\
    & \frac{\partial \conc}{\partial \eta} =  \kin \delfunc(\xi,\eta) = \kin \delfunc(\xi) ,& &\eta = 0, \xi \in \mathbb{R},
\end{align}
\end{subequations}
where we denote $\nabla_{\zeta 1}^2:= \frac{\partial^2 }{\partial \xi^2} + \frac{\partial^2 }{\partial \eta^2}$.  Using Neumann-Green's functions \cite{constantin2010green}, \eqref{zetaeqs} is solved by
\begin{equation}
    c = A_1 + \dfrac{\kin}{2 \pi}\log{(\eta^2 + \xi^2)}= A_1 + \dfrac{\kin}{\pi} \log{|\zeta_1|} = \dfrac{\kin}{\pi} \Re{\log(\zeta_1)},
\end{equation} 
where $A_1$ and $\kin$ are constants that will be determined from matching later. Hence, our Region \mbox{IV} solution is
\begin{equation}
    \conc \sim  A_1 + \frac{\kin}{\pi} \Re (\log (\sin(\pi \zbl))). \label{steadsolbound1}
\end{equation}

The analysis in Region \mbox{II} follows from that in Region \mbox{IV} by symmetry. The appropriate scaling in this case is
\begin{equation}
    \y = \chanL + \delta \ybu,
\end{equation}
and, using the notation $\zbu = \xb + i \ybu$, is solved by
\begin{equation}
    \conc \sim A_2 + \frac{\kout}{\pi} \Re(\log(\zeta_{2})) = A_2 + \frac{\kout}{\pi} \Re(\log(\sin(\pi \zbu))), \label{steadsolbound2}
\end{equation}
where $A_2$ and $\kout$ are determined from later matching. This concludes our specific analysis of all inner and boundary regions. It is useful to determine the limiting forms of \labelcref{steadsolbound2,steadsolbound1} as $Y_{i} \rightarrow \mp \infty$ and $|Z_{i}| \rightarrow 0{_\mp}$, $i \in \{1,2\}$, for later matching purposes. First, taking the limit as $Y_{i} \rightarrow \pm \infty$, we obtain
\begin{equation}
    \conc \sim \amatch \mp \kmatch \ybmatch - \frac{\kmatch}{\pi} \log{2}  \quad \text{as } Y_{i} \rightarrow \mp \infty \quad i \in \{1,2\}, \label{matchbl1too}
\end{equation}
which we will use to match into the outer regions. Next, taking the limit as $|Z_{i}| \rightarrow 0_{\mp}$ we obtain
\begin{equation}
    \conc \sim \amatch + \frac{\kmatch}{ \pi} \log{|\pi Z_{i}|} \quad \text{as } |Z_{i}| \rightarrow 0_{\mp} \quad i \in \{1,2\},  \label{matchblltoin}
\end{equation}
which we match to our inner region solutions.

\subsection{Effective coupling conditions for the long thin channel limit} \label{matchsteady}
Our final task for the steady problem is to match our solutions between each asymptotic region. This will generate the asymptotically correct connection formulae between the outer concentrations and concentration fluxes, which are the effective coupling conditions we have been seeking. We match from our outer regions inwards, allowing us to determine our unknown constants in each region entirely in terms of outer quantities. First, Taylor expanding the outer solutions about $\y = \pm \chanL$, we obtain
\begin{equation}
    \conc(\x,\y) = \conc(\x, \pm \chanL + \delta Y_{i}) \sim \conc\rvert_{\pm \chanL} + \delta Y_{i}\frac{\partial\conc}{\partial \y}\rvert_{\pm \chanL}  \hspace{1cm} i \in\{1,2\}. \label{matcho1}
\end{equation}
Matching our outer solutions, \eqref{matcho1}, to our Regions \mbox{II} and \mbox{IV} solutions in the limit $Y_{i} \rightarrow \mp \infty$, $i \in \{1,2\}$, \eqref{matchbl1too} we deduce that
\begin{subequations}
\label{outertoboundrelations}
\begin{alignat}{4}
\kin &= - \delta \frac{\partial\conc}{\partial \y}\rvert_{y=-\chanL}, \quad 
&& \kout = \delta \frac{\partial \conc}{\partial y}\rvert_{\y = \chanL}, \label{match2}\\
A_1 &= \conc\rvert_{y=-\chanL} - \frac{\delta}{\pi} \frac{\partial\conc}{\partial \y}\rvert_{y=-\chanL} \log{2}, \quad&& 
A_2 = \conc\rvert_{\y = \chanL} + \frac{\delta}{\pi}\frac{\partial \conc}{\partial y}\rvert_{\y = \chanL} \log{2}.
\end{alignat}
\end{subequations}
Similarly matching our Regions \mbox{III}a and \mbox{III}c solutions as $\zinmatch \rightarrow \mp \infty$, \labelcref{in3atobl,in3ctobl} to our Regions \mbox{II} and \mbox{IV} solutions in the limit $Z_{i} \rightarrow 0$, $i \in \{1,2\}$, \eqref{matchblltoin}, we obtain
\begin{subequations}
\label{boundtoinnerrelations}
\begin{alignat}{2}
&2\ina = \kin, \quad && 2 \atild = \kout,\label{mbin1}\\
&\inb = A_1 + \frac{\kin}{\pi} \log{2\eps}, \quad &&\btild = A_2  + \frac{\kout}{\pi}\log{2 \eps}. \label{mbin2}
\end{alignat}
\end{subequations}

Finally matching the solutions in Regions \mbox{III}a and \mbox{III}c, \labelcref{in3atoin3b,in3ctoin3b}, to the solution in Region \mbox{III}b, \eqref{steadsolin2}, we obtain relations between $\ina, \atild, \inb$ and $\btild$,

\begin{equation}
\alpha = - \frac{\ina}{\delta \eps}= \frac{\atild}{\delta \eps}, \quad \beta = \inb + \alpha \chanL + \frac{2 \ina}{\pi}(\log{2}-1) = \btild - \alpha \chanL + \frac{2 \atild}{\pi}(\log{2}-1). \label{innertochannelrelations}
\end{equation}

Substituting \eqref{outertoboundrelations} into \eqref{boundtoinnerrelations} we can write $a_{\pm},b_{\pm}$ in terms of the outer concentration, concentration flux, and membrane geometry parameters. Subsequently using \eqref{innertochannelrelations}, which gives us the relationship between $a_{\pm},b_{\pm}$, we arrive at our effective coupling conditions, directly relating the outer flux and concentrations on either side of the membrane via the classic permeability equation
\begin{subequations} \label{fullsteadcon}
 \begin{equation}
   \frac{\partial\conc}{\partial n}\rvert_{y = -L}= \frac{\partial \conc}{\partial n} \rvert_{y=L}=P_{\text{eff }}\left[\conc\right]^L_{-L}, \label{permeqstead}
\end{equation} 
\begin{equation}
    P_{\text{eff}} = \frac{1}{\frac{\chanL}{\eps} +  \frac{2\delta }{\pi}\left(\log{\frac{1}{8 \eps}}+ 1 \right)}. \label{steadperm}
\end{equation}
\end{subequations}
The effective coupling conditions \eqref{fullsteadcon} have the same functional form as \eqref{permeationeq}, but now with an explicit expression for the effective permeability $P_{\text{eff}}$, \eqref{steadperm}, as a function of the physical gap properties of length, width and separation distance. This allows us to directly understand how changing microscale geometry affects membrane permeability, and hence the concentration jump across the membrane. Continuity of flux is encoded in \eqref{permeqstead} (and \eqref{permeationeq}), which follows physically since flux is conserved throughout the inner regions. We validate \eqref{fullsteadcon} against numerical simulations of the full membrane geometry in \S \ref{numerics}.

\subsection{The $\mathcal{O}(1)$ aspect ratio limit} \label{o1limit}
We present the $\mathcal{O}(1)$ aspect ratio limit, where $\delta \eps = \mathcal{O}(\chanL$), in detail in Appendix \ref{o1details}, summarising the main results here. We introduce the channel aspect ratio $a = {\chanL}/{\delta \eps} = \mathcal{O}(1)$ for notational convenience. Briefly, the main change from our analysis above is that our inner regions \mbox{III}a, \mbox{III}b and \mbox{III}c coalesce into a single inner region encompassing the gap openings and length. We can solve this problem using a different complex variable mapping to transform our gap domain to the upper half-plane; namely the inverse of the conformal map
\begin{equation}
  Z= X + iY = a i + d\left\{ \omega \left(1-\dfrac{1}{\omega^2}\right)^{\frac{1}{2}} \left(1-\dfrac{\lambda^2}{\omega^2}\right)^{\frac{1}{2}} - \text{E}(\arcsin{\frac{1}{\omega},\lambda)} - \lambda  \text{E}(\arcsin{\frac{\lambda}{\omega}},\frac{1}{\lambda})\right \},  \label{aspectconfmap}
\end{equation}
where $\{X,Y\} = \{x,y\}/\delta\eps$ are our inner variables describing the new Region \mbox{III}, $\omega = u+iv$, $v >0$, is our mapping variable in the upper-half plane as in \eqref{schwarzmap}, and $\text{E}(\phi,k)$ is the incomplete elliptic integral of the second kind defined by
\begin{equation}
    \text{E}(\phi,k) =\int_{0}^{\sin{\phi}} \dfrac{\sqrt{1-k^2t^2}}{\sqrt{1-t^2}}\mathrm{d}t. \label{elliptic}
\end{equation}
As in \eqref{schwarzmap}, we choose the branch cuts for \labelcref{aspectconfmap} to lie outside the physical domain. The constant $d$ is defined as
\begin{equation}
    d = - \dfrac{1}{\left\{ \text{E}(\arcsin{\frac{1}{\lambda}}, \lambda) + \lambda \text{E}(\frac{\pi}{2}, \frac{1}{\lambda}) \right\}} \in \mathbb{R}, \label{o1const}
\end{equation}
where the constant $\lambda(a) > 1$ is defined implicitly through the relation
\begin{equation}
    2a = \left({\int}_{1}^{\lambda}\dfrac{\left(\lambda^2 - t^2 \right)^{\frac{1}{2}}}{\left(t^2-1\right)^{\frac{1}{2}}} \mathrm{d}t - \lambda {\int}_{\frac{1}{\lambda}}^{1}\dfrac{\left(t^2 - \frac{1}{\lambda^2}\right)^{\frac{1}{2}}}{\left(1-t^2\right)^{\frac{1}{2}}} \mathrm{d}t\right) \left({\int}_{0}^{1}\dfrac{\left(\lambda^2 - t^2\right)^{\frac{1}{2}}}{\left(1-t^2\right)^{\frac{1}{2}}} \mathrm{d}t - \lambda {\int}_{0}^{\frac{1}{\lambda}}\dfrac{\left(\frac{1}{\lambda^2}-t^2\right)^{\frac{1}{2}}}{\left(1-t^2\right)^{\frac{1}{2}}} \mathrm{d}t\right) ^{-1}, \label{lambdaimplicit}
\end{equation}
noting that $\lambda \rightarrow 1$ as $a \rightarrow 0$. Proceeding as before, we use the limiting behaviour of the solution and the map \eqref{aspectconfmap} to match into our boundary layer regions. This procedure yields our effective coupling conditions, which again take the form of the classic permeability equation \eqref{permeationeq}, but now $P_{\text{eff}}$ is given by
\begin{equation}
    P_{\text{eff}} = \dfrac{1}{\dfrac{\delta}{\pi}\log{\left(\dfrac{1}{4\eps^2 \pi^2 \lambda d^2}\right)}}. \label{aspectperm}
\end{equation}

\begin{figure}
    \begin{overpic}[scale=0.55, percent ,grid=false]{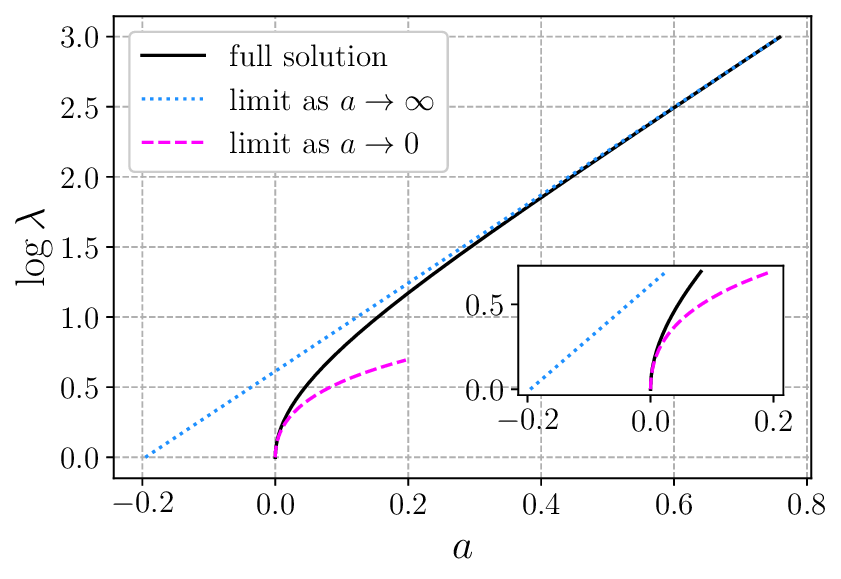}
    \put(2,70){(a)}
    \put(102,0){\includegraphics[scale=0.55]{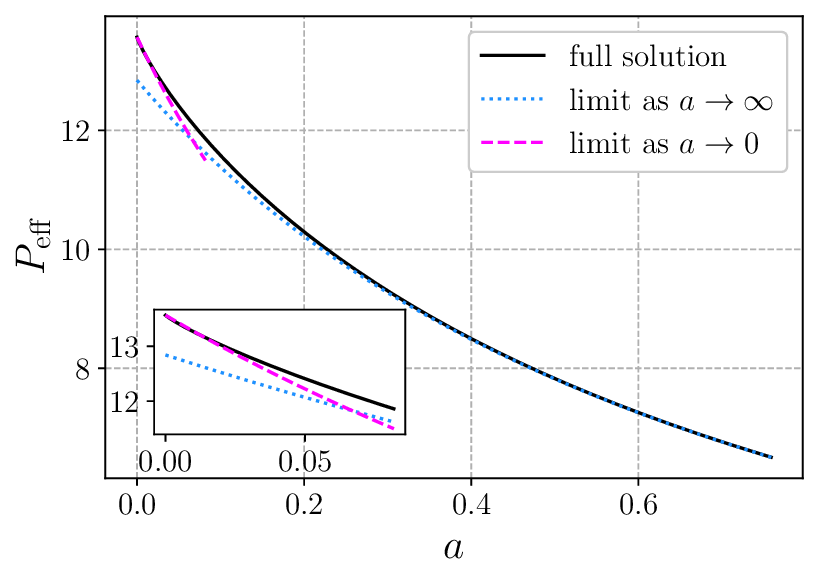}}
    \put(102,70){(b)}
    \end{overpic}
    \caption{(a) Aspect ratio $a={\chanL}/\delta \eps$ vs $\lambda$ as given by \eqref{lambdaimplicit}. We include the leading-order term in the expansion as $a \rightarrow 0$, \eqref{lamasmall},  (pink), alongside the leading order expansion as $a \rightarrow \infty$, \eqref{alamlargelim},  (blue). (b) Aspect ratio $a={\chanL}/{\delta\eps}$ vs $P_{\text{eff}}$ as given by \eqref{aspectperm}. We include the leading-order limiting forms as $a \rightarrow 0$, \eqref{smallalim}, (pink) and $a \rightarrow \infty$, \eqref{largealim} (blue). Inset plots show behaviour for small $a$.}
    \label{aspectvlamandperm}
\end{figure}

We now investigate how the effective permeability changes in this $O(1)$ aspect ratio limit, \eqref{aspectperm}, compared to the long thin limit, \eqref{steadperm}. To do this, we examine the dependence of $\lambda$ on $a$, written implicitly in \eqref{lambdaimplicit} and plotted in Figure \ref{aspectvlamandperm}(a). We see that $a \rightarrow 0$ corresponds to $\lambda \rightarrow 1$ and $a \rightarrow \infty$ corresponds to $\lambda \rightarrow \infty$. As such, we now examine the leading-order approximations of the integrals in \eqref{lambdaimplicit} in the limits $\lambda \rightarrow1$ and $\lambda \rightarrow \infty$. 


We first explore how a small aspect ratio affects the form of the effective permeability by taking the limit $\lambda \rightarrow 1$ (and therefore $a \rightarrow 0$) in \eqref{lambdaimplicit} to obtain
\begin{equation}
     a \sim \dfrac{\pi}{16}(\lambda-1)^2 \quad \text{as } a \to 0,\label{lamasmall}
\end{equation}
to leading order. Substituting \eqref{lamasmall} into \eqref{aspectperm} we deduce
\begin{equation}
    P_{\text{eff}} \sim \dfrac{1}{\dfrac{2 \delta}{\pi}\log{\dfrac{1}{\varepsilon \pi}} +\dfrac{16}{\pi^2} \delta a} \sim \dfrac{1}{\dfrac{2 \delta}{\pi}\log{\dfrac{1}{\varepsilon \pi}} }\quad \text{as } a \to 0, \label{smallalim}
\end{equation}
where we have used $d \sim -1/2 + ({\lambda-1})/{4}$ as $\lambda \rightarrow 1$ in \eqref{o1const}. We note that in this limit we require $\eps$ sufficiently small, $\eps \ll \exp(-1/\delta)$, to sustain a leading-order concentration difference across the membrane. For larger $\eps$ we find $P_{\text{eff}} \rightarrow \infty$ and we recover the results in \cite{zampogna_transport_2022}, where concentration is equal on either side of an infinitely thin membrane at leading order.
Similarly, we explore the large-aspect-ratio limit by taking $\lambda \rightarrow \infty$ (and therefore  $a \rightarrow \infty$) in \eqref{lambdaimplicit} to obtain
\begin{equation}
    a \sim \dfrac{1}{\pi}(\log{4\lambda}-2) \quad \text{as } a \to \infty. \label{alamlargelim}
\end{equation}
Substituting \eqref{alamlargelim} into \eqref{aspectperm} we now obtain
\begin{equation}
 P_{\text{eff}} \sim \dfrac{1}{\dfrac{2\delta}{\pi}\left( \log{\dfrac{1}{8 \varepsilon}} + 1 \right) + a \delta} \quad \text{as } a \to \infty,\label{largealim}   
\end{equation}
using $d \sim - {2}/{\pi \lambda}$ as $\lambda \rightarrow \infty$ in \eqref{o1const}.
Writing $a = \chanL/\delta \eps$, we find that the large-$a$ limit of the $O(1)$ aspect ratio permeability \eqref{largealim} recovers the effective permeability \eqref{steadperm} we derived separately from our long thin channel limit analysis. We note that the effective permeability in both small and large aspect ratio limits, \labelcref{smallalim,largealim}, depends on the membrane geometry through the parameters $a, \delta$ and $\eps$ with the same functional form, but with different pre-multiplying constants.

In Figure \ref{aspectvlamandperm}(b) we show how the effective permeability, $P_{\text{eff}}$, varies with the aspect ratio, $a = \chanL/\delta \eps$, including the full effective permeability in our $\mathcal{O}(1)$ aspect ratio limit, \eqref{aspectperm} and the two limiting forms \labelcref{smallalim,largealim}. We see that the large aspect ratio limit matches well even as the aspect ratio becomes fairly small, with good agreement for $a \gtrsim 0.25$, meaning that our simpler long thin channel limit performs well as an effective interface condition unless the membrane is much thinner than the width of the gaps.

\section{Time-dependent problem} \label{s4timedep}
We now return to the full unsteady problem. Importantly, the genuine unsteady nature of the problem for $t = \mathcal{O}(1)$ is only relevant for the upscaling procedure in the long thin channel limit. This is because the small lengthscales in the boundary layer and inner regions lead to quick equilibration in the $\mathcal{O}(1)$ aspect ratio limit, and so \eqref{permeationeq} and \eqref{aspectperm} would still hold, albeit in a quasi-steady manner. However, in the long thin channel limit, the channel length $\chanL$ can be large enough so that in Region \mbox{III}b transport across the channel is not quasi-steady. As we shall see, this introduces additional technicalities and emergent effects when deriving effective coupling conditions. Specifically, we find that the coupling conditions we derive gain memory properties, where the history of the system becomes important. Given that the only significant change between analysis in the time-dependent problem is in Region \mbox{III}b, we focus on this region below.

\subsection{Region \mbox{III}b}
As before, Region \mbox{III}b describes the channel, where $\x = \mathcal{O}(\delta \eps) $ and $\y = \mathcal{O}(\chanL)$.
We scale into this region using
\begin{equation}
    \x = \delta \eps \inonex\; \hspace{1cm} \y = -\chanL + 2\timey, \label{timescalings}
\end{equation}
where our $\y$ scaling is chosen for later notational convenience. Our channel is now defined by $\{ (\inonex,\timey): |\inonex| < 1, 0< \timey < \chanL \}$. Substituting the scalings \eqref{timescalings} into our unsteady system \eqref{nondimgoverning} for our channel region, we obtain
\begin{subequations}
\label{inner2governtime}
 \begin{align}
    (\delta \eps)^2 \frac{\partial \conc}{\partial \ndtime} &= \frac{\partial^2 \conc}{\partial \inonex^2} + \frac{(\delta \eps)^2}{4} \frac{\partial^2 \conc}{\partial \timey^2},& \hspace{1cm} &|\inonex|<1, \quad 0 < \timey< \chanL, \label{timeeq1}\\
    \frac{\partial \conc}{\partial \inonex} &= 0,& & |\inonex|=1,\quad 0 < \timey< \chanL.\label{timeeq2}
\end{align}  
\end{subequations}

In the same manner as in \S 3(a)\ref{channellength}, we deduce that the leading-order concentration is independent of $\inonex$, so now we have $\conc \sim \conc(\y, t)$, and hence the leading-order concentration satisfies
\begin{equation}
    \frac{\partial \conc}{\partial \ndtime} \sim \frac{1}{4} \frac{\partial^2 \conc}{\partial \timey^2}, \hspace{1cm} 0 < \timey < \chanL. \label{timechannel}
\end{equation}
We can write the general solution of \eqref{timechannel} as
\begin{equation}
    \conc(\timey, \ndtime) \sim \alpha(\ndtime) \timey + \beta(\ndtime) + \sum_{n=1}^{\infty} g_n(\ndtime) \sin{\lambda_n \timey}, \hspace{1cm} \lambda_n = \frac{n \pi}{\chanL}, \label{gensoltime}
\end{equation}
where the functions $\alpha(\ndtime)$ and $\beta(\ndtime)$ are the time-dependent analogues to the constants of integration in the steady problem \eqref{steadsolin2}, and will be determined via later matching with the other regions. As we shall see later, the functions $g_n(\ndtime)$ represent genuinely unsteady effects, and splitting the solution as in \eqref{gensoltime} generates helpful smoothness properties whereby the $\bar{y}$-derivative of the infinite sum in \eqref{gensoltime} is convergent.

Substituting \eqref{gensoltime} into \eqref{timechannel}, multiplying by $\sin{\lambda_m \timey}$, and integrating over $\timey$ in $(0, \chanL)$, we obtain the following ODE for each $g_n(t)$:
\begin{equation}
  \frac{d}{dt}{g}_n+ \frac{\lambda_n^2}{4} g_n =\frac{d}{dt}\left(   \frac{2 {\alpha}}{\lambda_n}(-1)^n + \frac{2 {\beta}}{\chanL \lambda_n}((-1)^n-1)\right). \label{ODEg}
\end{equation}
Solving the linear ODE \eqref{ODEg} using standard methods, we obtain the solution
\begin{equation}
    g_n(\ndtime) = \exp\left(-\frac{\lambda_n^2 \ndtime}{4}\right)g_n(0) + \exp\left(-\frac{\lambda_n^2 \ndtime}{4}\right)\int_{0}^{\ndtime} \exp\left(\frac{\lambda_n^2 \tau}{4}\right) \frac{d}{d\tau}\left (   \frac{2 {\alpha}(\tau)}{\lambda_n}(-1)^n + \frac{2{\beta}(\tau)}{\chanL \lambda_n}((-1)^n-1) \right) \; \mathrm{d} \tau. \label{gensolg}
\end{equation}
Reassuringly, $g_n(\ndtime) \rightarrow 0$ as $\ndtime \rightarrow \infty$ if $d \alpha/ dt, d \beta/dt \rightarrow 0$ as $\ndtime \rightarrow \infty$, from which we can recover the steady solution \eqref{fullsteadcon}.

Returning to the time-dependent solution \eqref{gensolg}, with initial condition represented by $\conc_{\text{init}}(\timey)$ in the channel length, and writing in terms of a Fourier series, we have
\begin{equation}
    g_n(0) = \frac{2\alpha(0)}{\lambda_n}(-1)^n + \frac{2 \beta(0)}{\chanL \lambda_n}((-1)^n-1) +  \frac{2}{L} \int_{0}^{\chanL} \conc_{\text{init}}(\timey)  \sin{\lambda_n \timey} \; \mathrm{d} \timey. \label{initialg}
\end{equation}
For brevity we will henceforth set $\conc_{\text{init}}(\timey)=0$, but it is straightforward to extend the following to general initial conditions if required. 

Our final task is to match our solutions to determine $\alpha(\ndtime)$ and $\beta(\ndtime)$ in terms of outer variables. This will give us the effective coupling conditions we seek by formally connecting the outer variables on either side of the membrane.

\subsection{Time-dependent effective coupling conditions} \label{timedepconditionssec}
As time only arises as an effective parameter in the boundary layer and Region \mbox{III}a and \mbox{III}c solutions, the matching between those regions is essentially the same as for the steady case. This generates the relations \eqref{outertoboundrelations} and \eqref{boundtoinnerrelations}, but with matching functions of time $\kappa_{1,2}$, $A_{1,2}$, $a_\pm$ and $b_{\pm}$ instead of parameters. The fundamental difference in the unsteady problem is the explicit time dependence in Region \mbox{III}b. This means that the matching between Region \mbox{III}b and its neighbouring Regions \mbox{III}a and \mbox{III}c needs to be considered separately, which we discuss below.

To match, it is helpful to expand \eqref{gensoltime}, our solution in Region \mbox{III}b, in terms of its derivatives. We expand \eqref{gensoltime} about $\timey = 0$ and  $\timey = \chanL$, using the Regions \mbox{III}a and \mbox{III}c scalings, $Y_{\pm}$, \labelcref{inreg1scal,inreg3scal}, to obtain
\begin{subequations}
\label{taylorL}
\begin{align}
  &\conc(\timey, \ndtime) \sim \conc(\timey=0; \ndtime) + \frac{\delta \eps}{2} \inoney \frac{\partial \conc}{\partial \timey}(\timey=0; \ndtime) \qquad \text{as } \timey \to 0,\\
  &\conc(\timey, \ndtime) \sim \conc(\timey=\chanL; \ndtime) + \frac{\delta \eps}{2} \inthreey \frac{\partial \conc}{\partial \timey}(\timey=\chanL; \ndtime) \qquad \text{as } \timey \to L.
  \end{align}
\end{subequations}
We will match into Regions \mbox{III}a and \mbox{III}c using the values of the Region \mbox{III}b concentrations and concentration fluxes at the Region \mbox{III}b endpoints.

Matching Regions \mbox{III}a and \mbox{III}c with their respective outer regions follows in a similar manner to \S \ref{matchsteady}. We can therefore expand our leading-order solutions in Regions \mbox{III}a and \mbox{III}c, \labelcref{steadsolin1,steadsolin3}, as they approach the channel entrance, to obtain
\begin{equation}
\conc \sim  \conc_{\mp} \pm \frac{\delta}{\pi} \frac{\partial \conc_{\mp}}{\partial y}\left(1  + \log{\frac{1}{8 \eps}} \right) + \frac{\delta}{2} \frac{\partial \conc_{\mp}}{\partial y} \inymatch, \hspace{1cm} \inymatch \rightarrow 0^{\mp}\label{matchtimeinner},
\end{equation}
where for notational convenience we have introduced the representations $\conc_{\pm} := \conc(y = \pm\chanL, \ndtime)$.
Comparing \eqref{matchtimeinner} with \eqref{taylorL}, we deduce the values of $\alpha(t)$ and $\beta(t)$ in \eqref{gensoltime}:
\begin{subequations}
\label{alpandbet}
\begin{align}
 &\alpha(\ndtime) = \frac{1}{\chanL}[\conc]_{-}^{+} - \frac{\delta}{\chanL\pi} \left(1 + \log{\frac{1}{8\eps}} \right)\left( \frac{\partial \conc_{+}}{\partial y} + \frac{\partial \conc_{-}}{\partial y}\right),\label{alpha}\\   
&\beta(\ndtime)= \concbelow +\frac{\delta}{\pi}\left(1 + \log{\frac{1}{8\eps}} \right) \frac{\partial \conc_{-}}{\partial y}.\label{beta}
\end{align}
\end{subequations}

We are now able to write our concentration in Region \mbox{III}b purely in terms of the membrane microstructure and outer concentration and flux using \eqref{alpandbet}. We do this using the concentration flux at the endpoints of our channel to obtain our coupling conditions connecting the concentration and flux on either side of the membrane. We start by differentiating \eqref{gensoltime} with respect to $\timey$ and evaluate for the flux at the endpoints of our channel using \eqref{gensolg}. We then match this with the flux from Regions \mbox{III}a and \mbox{III}c, as given by \eqref{matchtimeinner}, to obtain
    \begin{align}
    \label{fluxrelationstime}
            \dfrac{\partial \conc_{-}}{\partial y} = \eps \alpha(\ndtime) + \eps \sum_{n=1}^{\infty}\lambda_n g_n(\ndtime), \qquad
        \dfrac{\partial \conc_{+}}{\partial y} = \eps \alpha(\ndtime) + \eps \sum_{n=1}^{\infty}\lambda_n g_n(\ndtime)(-1)^n,
     \end{align}
To obtain our effective interface conditions we combine \labelcref{fluxrelationstime} using \eqref{alpandbet}, relating the average flux and flux jump across the interface. First, the difference between the expressions in \eqref{fluxrelationstime} yields
\begin{subequations} \label{fluxdiff}
\begin{equation}
 \dfrac{\partial \conc_{+}}{\partial y}-\dfrac{\partial \conc_{-}}{\partial y} = \eps \sum_{n \text{ odd}}  \exp\left(- \frac{\lambda_n^2 \ndtime}{4}\right) \left \{ - 2 \lambda_n g_n(0) +4 \int_{0}^{\ndtime} F_n\left(\tau, {\conc_{+}}+{\conc_{-}}, \frac{\partial \concabove}{\partial y}-  \frac{\partial \concbelow}{\partial y}\right) \mathrm{d} \tau \right \}, 
\end{equation}
\begin{equation}
    F_n \left(\tau, {\conc_{+}}+{\conc_{-}}, \frac{\partial \concabove}{\partial y}-  \frac{\partial \concbelow}{\partial y} \right) := \exp\left(\frac{\lambda_n^2 \tau}{4}\right)\frac{d}{d \tau}\left(\frac{\concabove + \concbelow}{\chanL} - \frac{\delta }{\chanL\pi}\left(\log{\frac{1}{8 \eps}}+ 1 \right) \left( \frac{\partial \concabove}{\partial y}-  \frac{\partial \concbelow}{\partial y}\right) \right).
\end{equation}
\end{subequations}
Importantly, \eqref{fluxdiff}  represents the first main result from the time-dependent analysis: the time-dependent flux jump across the interface. We note that the infinite sum vanishes and \eqref{fluxdiff} tends to the steady condition of flux continuity $[\partial \conc/\partial y]_{-}^{+} = 0$ for large time if $\conc_{\pm}$ and $\partial{\conc}_{\pm}/\partial y$ become independent of time as $\ndtime \rightarrow \infty$.

We obtain the final coupling condition, the average flux across the interface, by summing the expressions in \eqref{fluxrelationstime} to arrive at
\begin{subequations}\label{timeconcdiff}
\begin{equation}
    \frac{1}{2}\left(\frac{\partial \conc_{+}}{\partial y}+\frac{\partial \conc_{-}}{\partial y} \right)= P_{\text{eff}}[\conc]^{+}_{-}  +\chanL P_{\text{eff}}\sum_{n \text{ even}} \exp\left(- \frac{\lambda_n^2 \ndtime}{4}\right) \left[ \lambda_n g_n(0) 
  + 2 \int_{0}^{\ndtime} G_n\left(\tau, {\conc_{+}}-{\conc_{-}},\frac{\partial \concabove}{\partial y}+  \frac{\partial \concbelow}{\partial y}\right) \mathrm{d} \tau \right],
\end{equation}
 \begin{equation}
     G_n \left(\tau, {\conc_{+}}-{\conc_{-}}, \frac{\partial \concabove}{\partial y}+  \frac{\partial \concbelow}{\partial y}\right) := \exp\left(\frac{\lambda_n^2 \tau}{4}\right)\frac{d}{d\tau}\left(\frac{\concabove - \concbelow}{\chanL} - \frac{\delta }{\chanL\pi}\left(\log{\frac{1}{8 \eps}}+ 1 \right)\left(\frac{\partial \concabove}{\partial y}+  \frac{\partial \concbelow}{\partial y} \right) \right),
 \end{equation}
\end{subequations}
 where $P_{\text{eff}}$ is our steady effective permeability derived in the long thin channel limit, \eqref{steadperm}.
 Equation \eqref{timeconcdiff} represents the second main result from our time-dependent analysis. This closes the time-dependent problem. We again note that the infinite sum vanishes if $\conc_{\pm}$ and $\partial{\conc}_{\pm}/\partial y$ become independent of time as $\ndtime \rightarrow \infty$, and that in this scenario \eqref{timeconcdiff} would tend to the steady coupling condition \eqref{fullsteadcon} at large time. We therefore interpret \eqref{timeconcdiff} as our time-dependent effective permeability equation.

The effective unsteady coupling conditions, \labelcref{fluxdiff,timeconcdiff}, represent
the key results from our time-dependent analysis. Importantly, these effective coupling conditions have gained a memory property from the time integrals, which is not present in the steady problem. This can be interpreted as a quasi-delay time for information about the environment on one side of the channel to cross to the other side. The importance of this memory contribution depends on the channel geometry, and also the past averages and differences of the concentrations and fluxes, incorporated through the time integral terms. Reassuringly, this memory contribution decays in the large-time limit if the outer concentrations tend to a steady solution; that is, in this scenario the large-time limits of the sums vanish in \labelcref{fluxdiff,timeconcdiff} and we recover the steady coupling conditions \eqref{fullsteadcon}.

Finally, we also note that the infinite sums in \labelcref{fluxdiff,timeconcdiff} are slow to converge as $\ndtime \rightarrow 0^{+}$. To circumvent this numerical issue, we use an asymptotic Euler-Maclaurin approximation to convert our slowly converging sums into integrals. We present full details in Appendix \ref{EMappend} and note that, assuming vanishing initial conditions for simplicity, the key result of this analysis is that we can write our coupling conditions, \labelcref{fluxdiff,timeconcdiff}, in a simpler closed form as $t \to 0^{+}$:
\begin{subequations}\label{EMConditions}
    \begin{align}
        \frac{\partial \conc_{+}}{\partial y}-\frac{\partial \conc_{-}}{\partial y} &\sim 4\eps \chanL \sqrt{\frac{t}{\pi}}\frac{\mathrm{d}}{\mathrm{d}t}\left[\frac{\concabove + \concbelow}{\chanL} - \frac{\delta }{\chanL\pi}\left(\log{\frac{1}{8 \eps}}+ 1 \right) \left( \frac{\partial \concabove}{\partial y}-\frac{\partial \concbelow}{\partial y}\right) \right],  \label{EM1}\\
    \frac{1}{2}\left(\frac{\partial \conc_{+}}{\partial y}+\frac{\partial \conc_{-}}{\partial y} \right) &\sim P_{\text{eff}}[\conc]^{+}_{-} + \chanL^2 P_{\text{eff}} \sqrt{\frac{t}{\pi}} \frac{\mathrm{d}}{\mathrm{d}t}\left[\frac{\concabove - \concbelow}{\chanL} - \frac{\delta }{\chanL\pi}\left(\log{\frac{1}{8 \eps}}+ 1 \right)\left(\frac{\partial \concabove}{\partial y}+  \frac{\partial \concbelow}{\partial y} \right) \right]. \label{EM2}
    \end{align}
\end{subequations}

\section{Numerical validation} \label{numerics}
We now compare simulations of a section of the full membrane problem against simulations with our effective membrane conditions, using the open source finite element python package FEniCSx, \cite{barrata2023dolfinx,scroggs2022basix,scroggs2022construction,AlnaesEtal2014} with meshes generated using the finite element mesh generator Gmsh \cite{geuzaine2009gmsh}. We take our full domain to have height and width $2R = 5$ in all simulations. In each of our simulations we impose no flux boundary conditions on the side walls, and Dirichlet boundary conditions on the top and bottom walls of $c=0$ and $c=1$ respectively. We validate the numerical method in Appendix \ref{appendix:numerical}.

\subsection{Steady simulations}
Our simple Neumann and Dirichlet boundary conditions yield an analytic solution to our steady effective problem. Representing our effective permeability as $P_{\text{eff}}$ in both the long thin channel limit, \eqref{steadperm}, and the $\mathcal{O}(1)$ aspect ratio limit, \eqref{aspectperm}, we deduce
\begin{equation}
c \sim
    \begin{cases}
        1 - \dfrac{ P_{\text{eff }} y}{1+ 2 P_{\text{eff}}(R-L)}  \hspace{2.2cm} y < R - L,\\
        \dfrac{P_{\text{eff}}(y - 2R)}{1+ 2 P_{\text{eff}}(R-L)} \hspace{2.8cm} y> R+L,
        \end{cases} \label{analyticalsol}
\end{equation}
where $\chanL$ is replaced with $a \delta \eps$ in the $\mathcal{O}(1)$ aspect ratio limit. 

We focus here on the long thin channel limit (similar results and simulations are obtained in the $\mathcal{O}(1)$ aspect ratio limit).
To evaluate the accuracy of our coupling conditions, \eqref{permeqstead}, we take $2L =0.5$ and vary the number and width of channels between simulations to explore different membrane permeabilities, effectively changing $\delta$ and $\eps$. 

\begin{figure}[t]
\begin{overpic}[abs,unit=1mm,scale=0.3,grid=false]{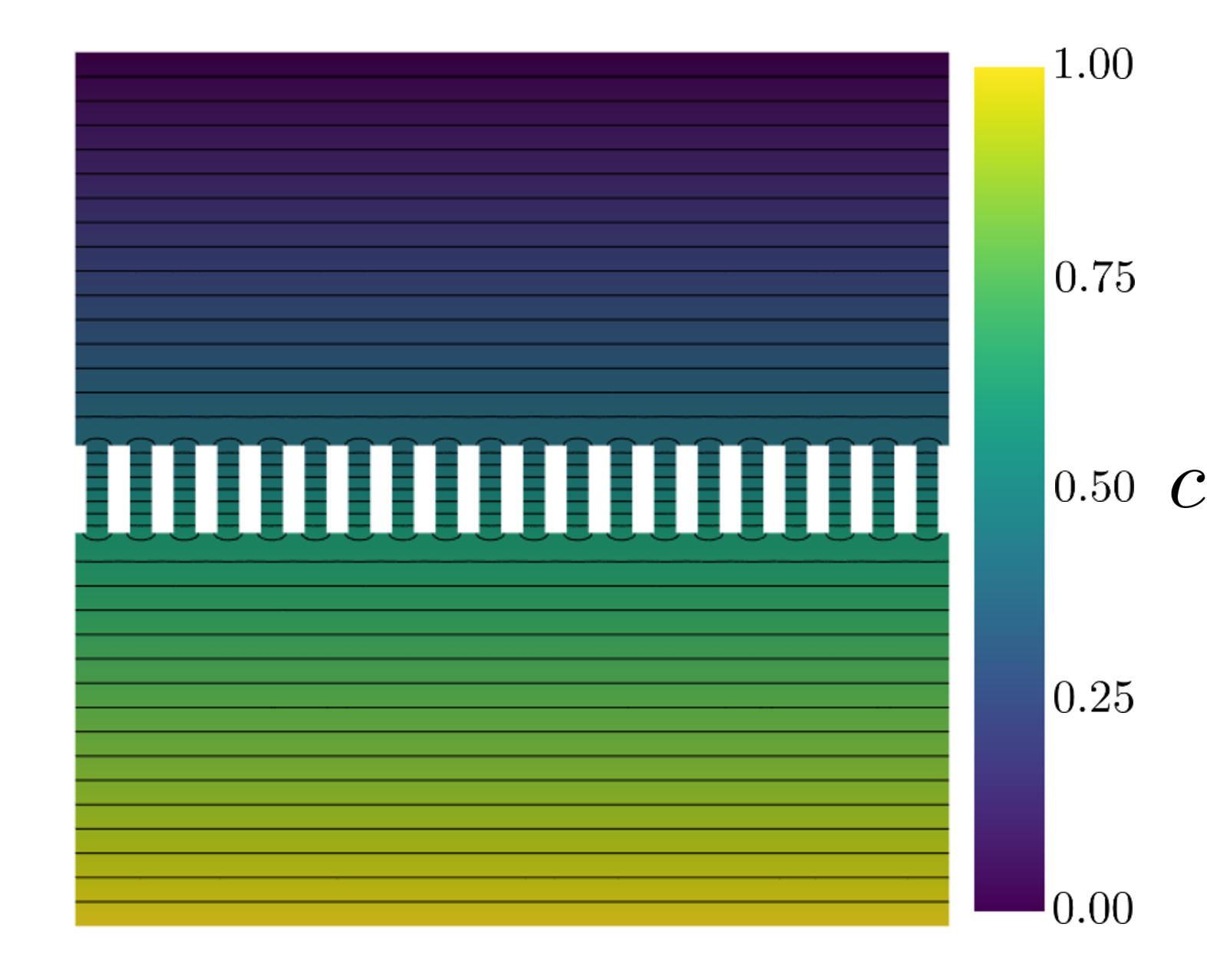}
\put(0,65){(a)}
\put(85,0){\includegraphics[scale=0.3]{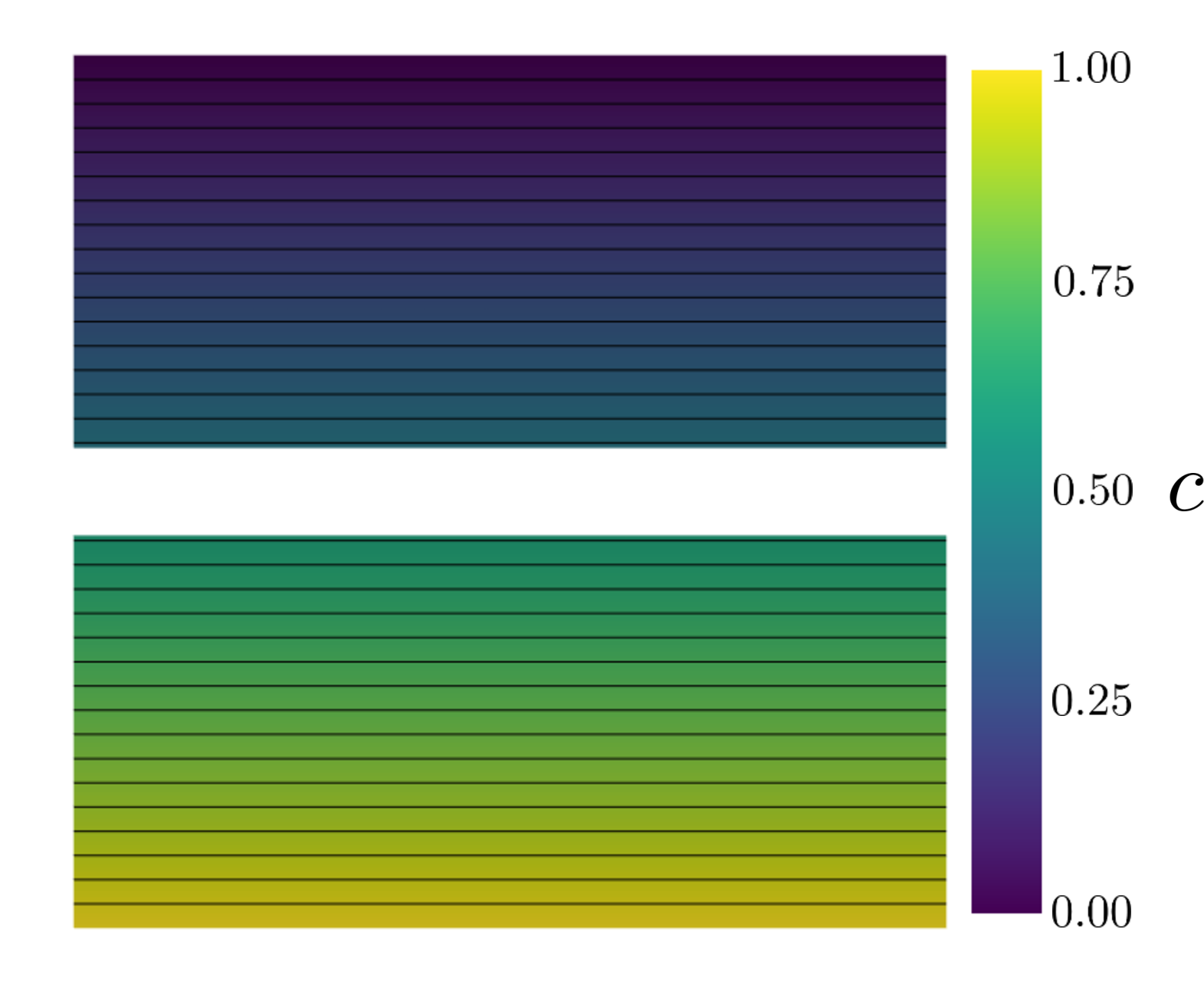}}
\put(83,65){(b)}
\end{overpic}
    \caption{Concentration profile for a membrane with effective permeability $P_{eff} =0.953$ and microscale parameters $\delta = 0.25$ $\eps=0.25$, for (a) a numerical simulation of the full membrane geometry, (b) the analytical solution with our effective interface condition \eqref{permeqstead}. Contours denote separations of 0.025. }
    \label{steadlonghighperm}
\end{figure}

\begin{figure}[t]
\begin{overpic}[abs,unit=1mm,scale=0.3,grid=false]{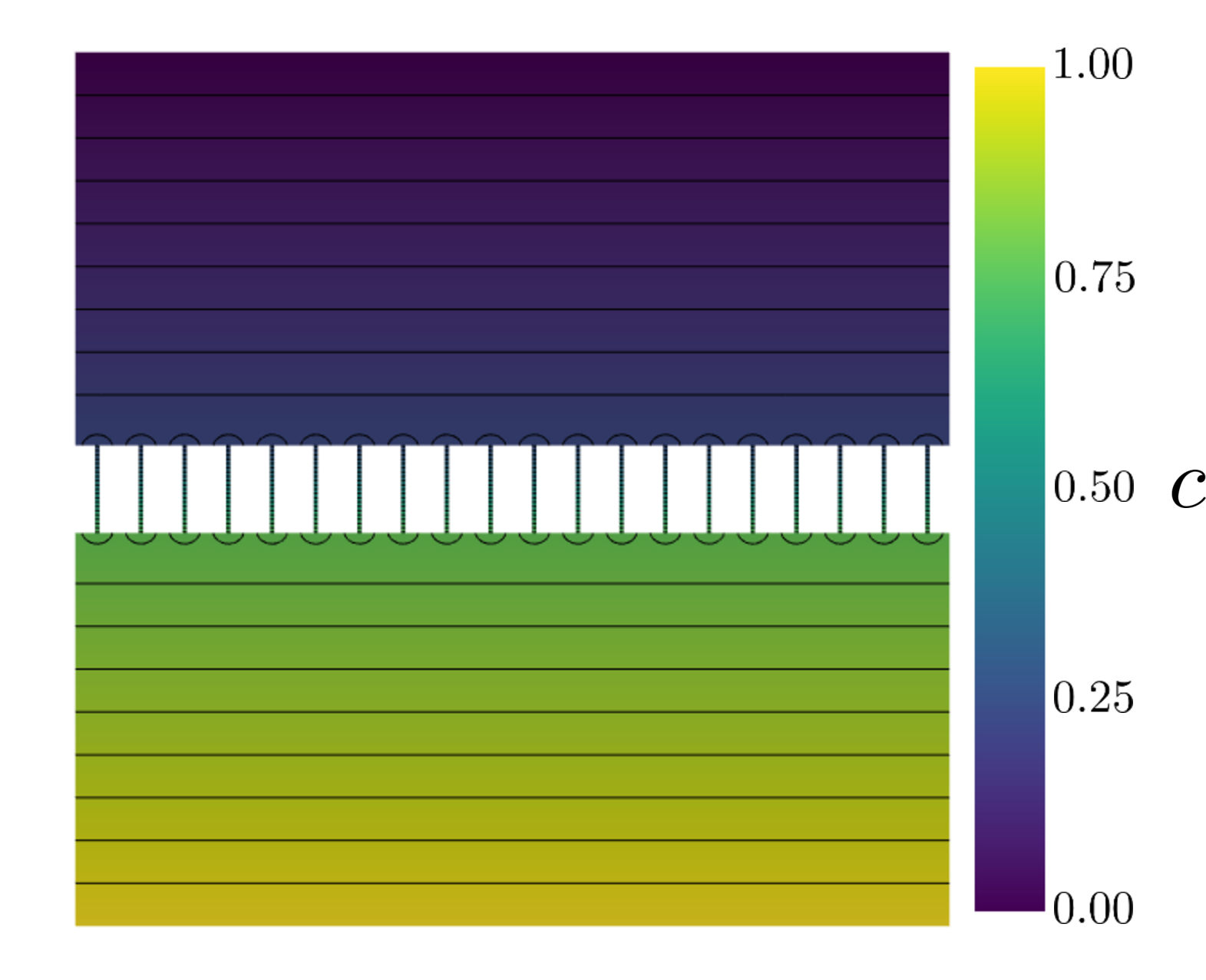}
\put(0,65){(a)}
\put(85,0){\includegraphics[scale=0.3]{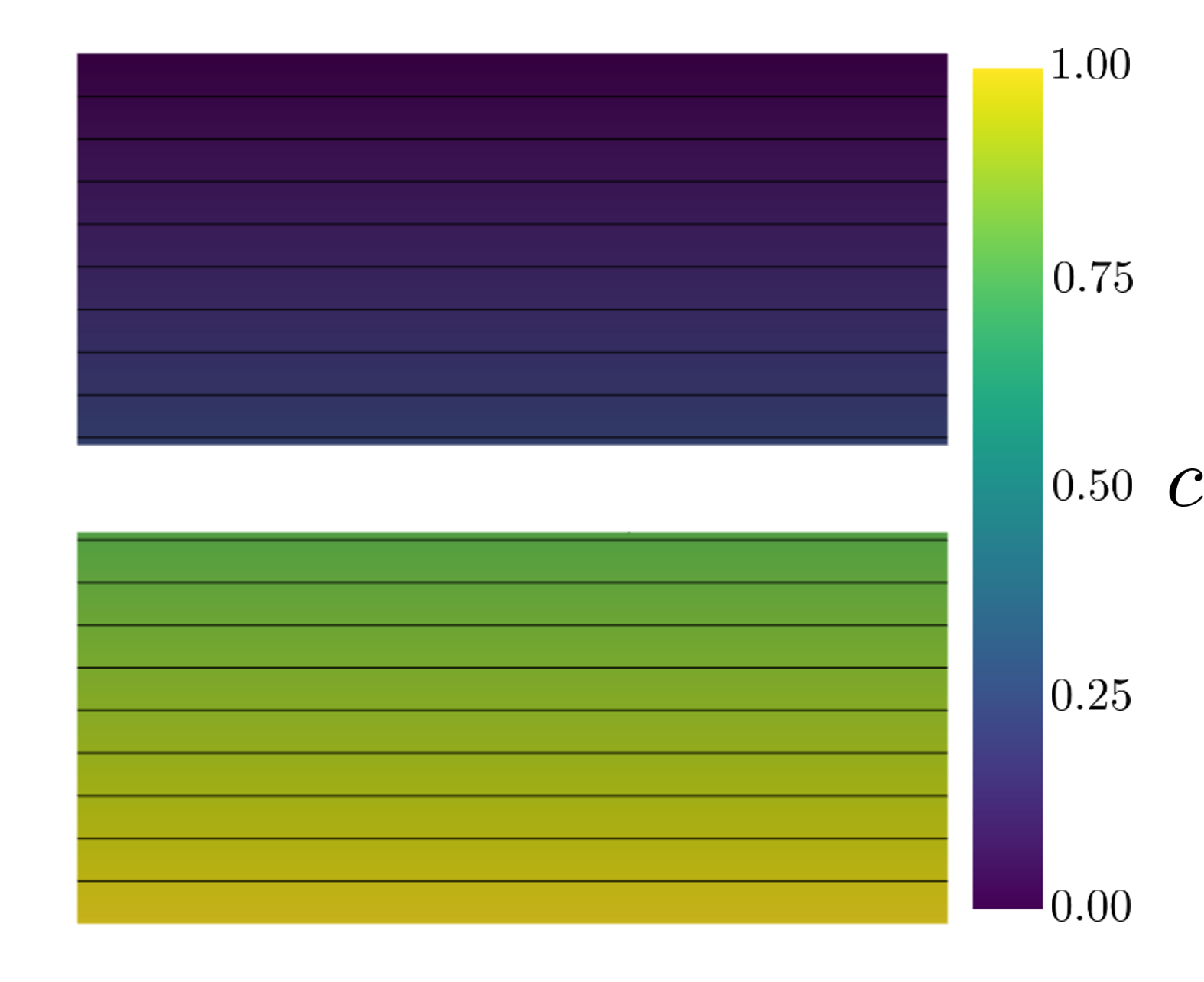}}
\put(83,65){(b)}
\end{overpic}
    \caption{Concentration profile for a membrane with effective permeability $P_{eff} =0.189$ and microscale parameters $\delta = 0.25$, $\eps = 0.05$ for (a) a numerical simulation of the full membrane geometry, (b) the analytical solution with our effective interface condition \eqref{steadperm}. Contours denote separations of 0.025.}
    \label{steadlonglowperm}
\end{figure}

\begin{figure}
    \begin{overpic}[scale=0.55,percent ,grid=false]{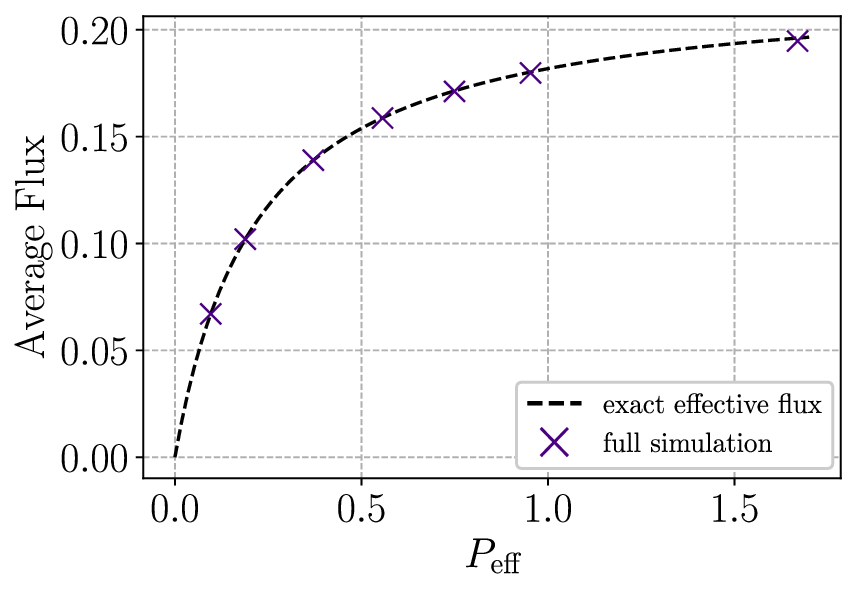}
    \put(0,74){(a)}
    \put(105,0){\includegraphics[scale=0.55]{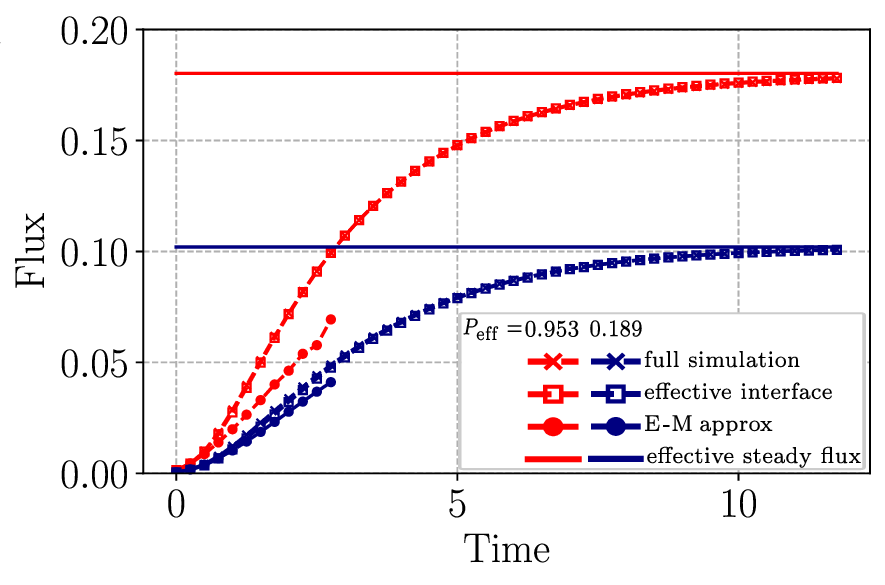}}
    \put(103,74){(b)}
    \end{overpic}
    \caption{(a) Average flux vs effective permeability in the steady long thin channel problem, \eqref{steadperm}. The black dashed line shows the analytical flux calculated from \eqref{analyticalsol}, with the full simulation added as purple crosses.
    (b) Average flux vs time in the unsteady problem for two different membrane geometries with steady effective permeabilities of 0.189 (blue) and 0.953 (red). We plot the flux for a numerical simulation of the full geometry (crosses), a numerical simulation of the effective interface condition with the infinite sum truncated at 30 terms, \labelcref{fluxdiff,timeconcdiff}, (squares) and a numerical simulation of the Euler-Maclaurin condition, \eqref{EMConditions}, up to $t=2.75$ (circles). The solid lines give the steady effective flux of the membrane geometry simulated calculated from \eqref{analyticalsol}. In both plots flux is calculated over a horizontal line away from the membrane.}
    \label{fluxvpermandtime}
\end{figure}

We obtain good visual agreement between the analytical solutions of our effective interface problem \eqref{analyticalsol} and simulations of transport in the full geometry, both at higher permeability (Figure \ref{steadlonghighperm}, $\delta =0.25$, $\eps =0.25$) and lower permeability (Figure \ref{steadlonglowperm}, $\delta =0.25$, $\eps =0.05$). We show the absolute value of the error between the analytical solution of our effective problem and the full simulation in Figure \ref{absdiff}in Appendix \ref{app:error}. We plot the flux across a horizontal line away from the membrane in simulations of the full problem alongside our analytical flux calculated from \eqref{analyticalsol} as we vary permeability in Figure \ref{fluxvpermandtime}(a). We see good agreement for a range of permeabilities, including for larger permeabilities. For example, the flux predicted by our coupling conditions matches the flux in the full numerical simulation to 99.4\% for a membrane with $P_{\text{eff}} = 1.5$, see Figure \ref{relfluxerrors}(a) in Appendix \ref{app:error}.

\subsection{Time-dependent simulations}
To evaluate the accuracy of our time-dependent coupling conditions \eqref{fluxdiff} and \eqref{timeconcdiff}, we carry out our time-dependent simulations using the inbuilt PETSc Krylov solver GMRES, with an LU preconditioner, using a timestep of $\Delta t = 0.25$ (timestep convergence is discussed in Appendix \ref{appendix:numerical} and shown in Figure \ref{numconv}(b)). We plot average flux over a horizontal line away from the membrane vs time in Figure \ref{fluxvpermandtime}(b) for simulations of the full membrane geometry, simulations using our effective interface conditions, \labelcref{fluxdiff,timeconcdiff}, and simulations using our Euler-Maclaurin early-time approximations for the interface condition, \labelcref{EM1,EM2}, for two different membrane geometries with effective permeabilities $P_{\text{eff}} = 0.189$ (blue) and $P_{\text{eff}} = 0.953$ (red). We see good agreement between all three simulations in both membranes, noting that we have simulated the Euler-Maclaurin approximations to $t=2.75$, beyond which they start to deviate more significantly from the other two simulations. We see convergence towards the steady flux value calculated from \eqref{steadperm} and \ref{analyticalsol}. We quantify the accuracy of the average flux in Figure \ref{relfluxerrors}(b) in Appendix \ref{app:error}.

\section{Permeability regimes and implications} \label{permregimes}
The straightforward nature of our derived steady coupling conditions in the long thin channel limit, \eqref{permeqstead} and \eqref{steadperm}, means that we can explicitly evaluate the contributions of the microstructure on the resulting macroscale membrane permeability. Our effective permeability coefficient \eqref{steadperm} depends explicitly on the microscale membrane geometry through the nondimensional parameters, $\chanL$, $\delta$ and $\eps$, representing the channel length, spacing and the ratio of channel width to spacing respectively.

We can write our permeability, \eqref{steadperm}, as
\begin{align}
P_{\text{eff}} \sim \dfrac{1}{\alpha_1 + \alpha_2}, \quad \text{with   } \alpha_1 = \dfrac{L}{\eps} \text{   and   } \alpha_2 = \dfrac{2 \delta}{\pi}\log \dfrac{1}{8\eps},
\end{align}
where $\alpha_1$ represents channel length effects and $\alpha_2$ represents the combined effects of the channel opening and interactions between neighbouring channels. With this notation, we can straightforwardly explore the dominant contributions of each physical effect by the relative sizes of our parameter groupings, as shown in Figure \ref{fig:permeability}. In each regime in Figure \ref{fig:permeability} our effective membrane condition will take the form of \eqref{permeqstead}, but with the effective permeability, \eqref{steadperm}, taking a different form depending on the limiting factor inhibiting transport. A higher permeability coefficient means the membrane is offering less resistance to transport whereas a lower permeability coefficient means more resistance to transport.\\ \\
\noindent \textbf{Regimes 1 and 2} are shown in green in the lower-right-hand half of Figure \ref{fig:permeability}. Here, transport is limited by channel length effects, with $\alpha_1 \gg \alpha_2$, and we can write the effective permeability as $P_{\text{eff}} \sim 1/\alpha_1 = L/\eps$. In these regimes, the effective permeability can be both large (in Regime 1 if $L \ll \eps$) and small (in Regime 2 if $L \gg \eps$), depending on the relative size of channel length compared to the ratio of channel width to spacing.
\\ \\
\noindent \textbf{Regimes 3 and 4} are shown in purple in the upper-left-hand half of Figure \ref{fig:permeability}. Here, the limiting factor in transport is channel opening and pore-pore interaction effects, with $\alpha_2 \gg \alpha_1$ and so we write the effective permeability as $P_{\text{eff}} \sim 1/\alpha_2  = \pi/(-2 \delta \log 8\eps)$. Here, we can also have both small and large effective permeabilities, depending on the relative sizes of channel width to channel separation. Specifically, the effective permeability is small in Regime 3 (where $8 \eps \ll \exp\left(-\pi/2 \delta\right)$), and it is large in Regime 4 (where $8 \eps \gg \exp\left(-\pi/2 \delta\right)$). This difference in system behaviour depends on the size of $\eps$ compared to an exponentially small function of $\delta \ll 1$. This critical behaviour when $\eps = \mathcal{O}(\exp\left(-\pi/2 \delta\right))$ agrees with the 2D critical scalings obtained in \cite{del_vecchio_thick_1987}, in which the authors also rigorously prove the form of the limiting problem for the thick Neumann sieve in $n$ dimensions for $n \geqslant 3$, and \cite{chapman_mathematics_2015, hewett2016homogenized}, in which the authors derive several models to describe shielding effects of the Faraday cage, including a homogenised continuum model.
\\ \\
\noindent \textbf{Regime 5} arises in the top left of Figure \ref{fig:permeability}, and is shaded grey. Here our long thin channel analysis technically breaks down as $\delta \eps = O(\chanL)$ and the analysis of our $\mathcal{O}(1)$ aspect ratio limit holds. However, from Figure \ref{aspectvlamandperm}, we see that our long thin channel limit still provides a good estimate of the effective permeability as we move into Regime 5, therefore we expect transport to be limited by channel opening and pore-pore interaction effects as in Regimes 3 and 4.

\subsection{Applications to bacterial membranes}

\begin{figure}[t]
\centering
    \includegraphics[width= 0.8\textwidth]{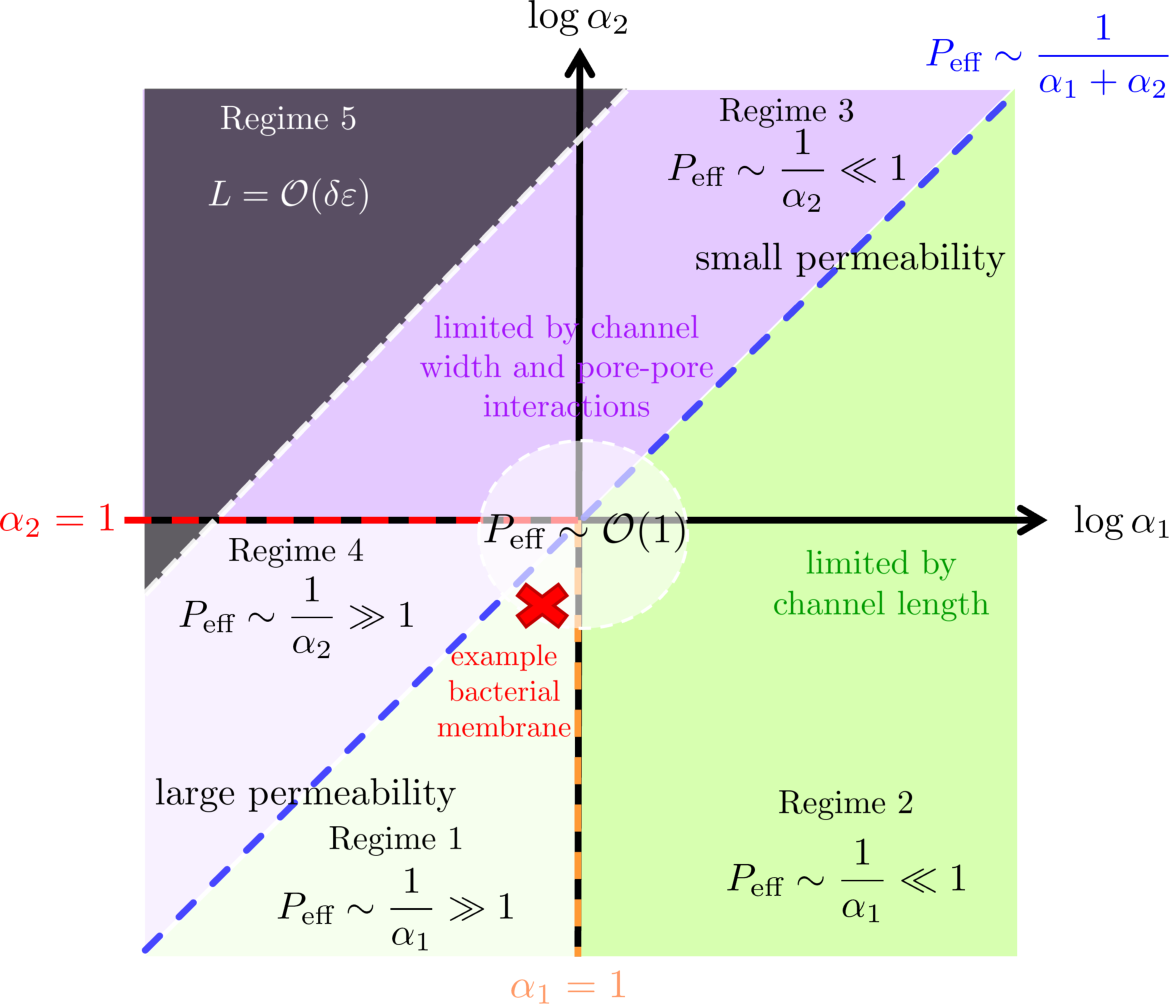}
   \captionof{figure}{Schematic showing how different parameter regimes of our microscale geometry affect the effective membrane permeability, \eqref{steadperm}, in the steady long thin channel limit. We group our parameters into $\alpha_1 = {L}/{\eps}$ and $\alpha_2 = {2 \delta}/{\pi}\log{\left({1}/{8\eps} \right)}$, where $L$ is the nondimensional channel length, $\delta$ the channel spacing and $\eps$ the ratio of channel width to channel spacing. The white dotted line is given by $\log{\alpha_2} = \log{\alpha_1} + \log{\left({2}/{\pi}\log{{1}/{8 \eps}}\right)}$ and marks the boundary in parameter space in which we cross into the $\mathcal{O}(1)$ aspect ratio limit. The dotted circle in the centre represents the scalings giving an $\mathcal{O}(1)$ effective permeability. When $\chanL = \mathcal{O}(\delta \eps)$ the white dotted line coalesces with the blue dashed line. Above the blue dashed line (in the purple region) permeability is dominated by channel width and pore-pore interaction effects and below this line (in the green region) it is dominated by channel length effects. Taking $\dimR=360\;$nm, $\dimsep=10\;$nm, $\dimwidth=0.325\;$nm and $2\dimL = 12\;$nm, from Table \ref{tab:par} we mark an example bacterial membrane by a red cross with $(\log{\alpha_1},\log{\alpha_2}) = (-0.65,-3.73)$.}
   \label{fig:permeability}
\end{figure}

Using the parameter values in Table \ref{tab:par} we find a range of $-7 \lesssim \log{\alpha_1} \lesssim -0.6$ and $\log{\alpha_2} \lesssim -3.6$, with permeabilities mostly found in Regimes 1, implying permeability in bacterial membranes is likely to be largely dominated by channel length effects. Taking our example parameters from before of $\dimR=360\;$nm, $\dimsep=10\;$nm, $\dimwidth=0.325\;$nm and $2\dimL = 12\;$nm, we obtain $\log{\alpha_1} = -0.65$ and $\log{\alpha_2} = -3.73$ placing us in a regime where permeability is $\mathcal{O}(1)$ and dominated by channel length effects, with $P_{\text{eff}} \approx 1.9$. We mark this by a red cross on Figure \ref{fig:permeability}, noting that these parameters are from a range of gram-negative bacteria and the membrane geometry of specific bacteria could lead to larger or smaller permeabilities with a potentially different limiting factor, which can be calculated using \ref{steadperm}. 

We predict that transport is limited by channel length effects when $\log{\alpha_2}> \log{\alpha_1}$. In dimensional parameters (defined in \eqref{nondiming}), this is equivalent to $\dimL  < 2 \dimwidth/\pi \log{(\dimsep/8\dimwidth)}$.
Using this we can predict how changes to the membrane microstructure could alter the limiting effect on transport. It is known that bacterial cells downregulate expression of porin genes in response to environmental stresses, allowing them to, for example, reduce intracellular antibiotic concentrations \cite{vergalli2020porins}. This effectively increases $\dimsep$ which could push permeability into channel width and entrance limited regimes. Similarly, mutant strains that show higher resistance to antibiotics are also found to express porins with reduced diameters, decreasing permeability by reducing $\dimwidth$ which can also shift behaviour into channel width and entrance limited regimes \cite{prajapati_how_2021}.


To complete our interpretation we return to the full steady permeability equation, converting \eqref{fullsteadcon} to dimensional variables using the transformations \eqref{nondiming} and the parameter relations \eqref{nondimpars}, obtaining
\begin{equation}
 \dimdif \frac{\partial \dimc}{\partial \dimy} \rvert_{\dimy=\dimL} = \dimdif\frac{\partial \dimc}{\partial \dimy}\rvert_{\dimy=-\dimL} =  \frac{\dimwidth \dimdif}{\dimsep \left(\dimL + \frac{2 \dimwidth}{\pi}\left[ \log{\frac{\dimsep}{8\dimwidth}} + 1\right] \right)} [\dimc]^{\dimL}_{-\dimL}, \label{dimfluxcon}
\end{equation}
where we have multiplied through by the diffusivity $\dimdif$ to turn the first two terms into genuine diffusive fluxes. We can write the effective coupling conditions \eqref{dimfluxcon} independently of $\dimdif$ due to their steady nature, but the unsteady conditions will have a genuine $\dimdif$ dependence. The dimensional formulation \eqref{dimfluxcon} allows us to directly consider the implications of the physical microscale geometry. Writing our channel length as $\dimfulll = 2\dimL$ and our channel width as $\dimfullwidth=2\dimwidth$, our dimensional permeability coefficient is given by
\begin{equation}
    \dimperm = \dfrac{\dimfullwidth}{\dimsep\left(\dimfulll + \dfrac{2\dimfullwidth}{\pi}\left[ \log{\dfrac{\dimsep}{4\dimfullwidth}} + 1\right] \right)}. \label{dimpermstead}
\end{equation} It is instructive to compare our derived effective permeability \eqref{dimpermstead} to the constitutive permeability one might pose assuming there are no interactions between porins. That is, to compare \eqref{dimpermstead} to the density of porins multiplied by the flux $J$ through a single isolated porin, calculated in \cite{prajapati_how_2021} to be
\begin{equation}
    J = \frac{A^{*} \dimdif}{\dimfulll} \Delta \dimc.\label{prajsingle}
\end{equation}
In \eqref{prajsingle}, $A^{*}$ represents the cross-sectional area of the porin (cf $\dimfullwidth$ here), $\dimfulll$ represents the length of the porin, $\dimdif$ is the diffusivity of $\dimc $, and $\Delta \dimc$ represents the difference in concentration at either end of the porin. If each porin has an independent effect on the flux, the total effect of all the porins on the macroscale membrane flux is
\begin{equation}
  \text{density} \times J = \dfrac{1}{\dimsep } \times \frac{A^{*} \dimdif}{\dimfulll} \Delta \dimc. \label{prajdens} 
\end{equation}
Our result \eqref{dimpermstead} takes the same form as \eqref{prajdens} in the limit in which permeability is dominated by porin length effects. That is, in Regimes 1 and 2 (Figure \ref{fig:permeability}) where $\chanL/\eps \gg (-2 \delta/\pi) \log 8\eps$, \eqref{dimfluxcon} reduces to 
\begin{equation}
\dimdif \frac{\partial \dimc}{\partial \dimy} \approx \frac{1}{\dimsep}\frac{\dimfullwidth \dimdif}{\dimfulll}[\dimc]^{\dimL}_{-\dimL}. \label{diminlim}
\end{equation}
where the right-hand sides of both \eqref{diminlim} and \eqref{prajdens} follow the same structure:
\begin{equation}
 \text{density}\times\frac{\text{channel cross-sectional \enquote{area}} \times \text{diffusivity}}{\text{channel length}}\times\text{concentration difference}.   
\end{equation}
We can understand this limit of our permeability equation to represent the case where there are no interactions between porins and each contributes individually to the overall flux. We can therefore interpret  $2\dimfullwidth/\pi \left[\log{\dimsep/4\dimfullwidth}+ 1\right]$, the term remaining in \eqref{dimpermstead}, as encoding the effect of the interaction between porins on membrane permeability.

\subsection{Analogy to resistivity}
Since the permeability $\dimperm$ is the reciprocal of the effective resistance imposed by the membrane, $R_{\text{tot}}^{*}$, we can use our explicit result \eqref{dimpermstead} to assess the individual contributions of microscale geometry to the effective resistance. This allows us to draw analogies with the resistance offered by resistors in an electrical circuit. Specifically, using the relation $\dimperm = 1/R_{\text{tot}}^{*}$ and \eqref{dimpermstead}, we can write
\begin{equation}
    \dfrac{R_{\text{tot}}^{*}}{\dimsep} = \dfrac{\dimfulll}{\dimfullwidth} + \dfrac{2}{\pi} + \dfrac{2}{\pi}\log{\dfrac{1}{4\dimfullwidth}}, \label{rtot}
\end{equation}
where the left-hand side represents the effective resistance multiplied by the porin density $1/\dimsep$. Equation \eqref{rtot} is analogous to the total resistance of an electrical circuit, given in terms of the resistance of resistors in series. With this analogy, we consider the contributions to total resistance that the concentration transport has to overcome in order to cross the membrane. The first term on the right-hand side of \eqref{rtot} relates to the resistance of the channel length, the second to a constant resistance (purely affected by the density of porins) and the third to the resistance offered by the porin openings. The interplay between how these various aspects of the microscale geometry affect permeability (and membrane resistance) is encoded in \eqref{dimpermstead} and \eqref{rtot}. Specifically we see that, as one might expect physically, thicker membranes (larger $\dimfulll$), narrower channels (smaller $\dimfullwidth$) and larger channel spacing (larger $\dimsep$) all lead to increased resistance and therefore decreased membrane permeability. Our results quantify the relative importance of these effects explicitly in terms of the membrane geometry.

\section{Discussion} \label{discuss}
In this paper we systematically derived effective coupling conditions for concentrations of solutes diffusing through narrow gaps in an otherwise impermeable membrane, allowing us to couple the concentration on either side of the membrane. We explored the limit in which membrane thickness is much greater than gap width, which we called the long thin channel limit, and the limit in which membrane thickness is of similar size to gap width, which we referred to as the $\mathcal{O}(1)$ aspect ratio limit. In the steady problem, the effective coupling conditions for both limits took the form of an effective permeability condition, \eqref{permeationeq}, written in terms of the concentration difference across the membrane and an effective permeability coefficient given by \eqref{steadperm} and \eqref{aspectperm} in the long thin channel and $\mathcal{O}(1)$ aspect ratio limits, respectively. This coefficient depends explicitly on the microscale geometry of the membrane, through the nondimensional parameters, $\chanL$, $\delta$ and $\eps$, representing the gap length, spacing and the ratio of gap width to spacing respectively. Using this effective permeability coefficient we were able to explore how changes in membrane microstructure alter the resulting permeability of the membrane, and explore various sublimits of the problem to determine the limiting factors affecting transport, and resulting size of the effective permeability. These results are summarised schematically in Figure \ref{fig:permeability}.

When considering the full time-dependent problem, the added complexity of the time it takes the solute to diffuse across the channel is captured within the infinite sums in our coupling conditions \labelcref{fluxdiff,timeconcdiff}. We see that the coupling conditions gain a memory property encoded through the time integrals in \labelcref{fluxdiff,timeconcdiff}. Provided that the outer concentration tends to a steady solution in the long time limit, i.e. $\dot{\conc}_{\pm} \rightarrow 0$, we find that the memory contribution decays and we recover the steady coupling conditions. Care needs to be taken evaluating the integrals and infinite sums in the limit $t \rightarrow 0^{+}$, where the sums are slow to converge. Here we approximate the infinite sums using an Euler-Maclaurin approximation to obtain simpler effective coupling conditions, as given in \eqref{EMConditions} and derived in Appendix \ref{EMappend}. This early-time approximation gives us insight into how the time-dependent conditions scale with $t$ and allow for faster simulations at early times, removing the need to include many terms in the infinite sums of \labelcref{fluxdiff,timeconcdiff}. For both the steady and time-dependent problems, we validated the derived effective coupling conditions against simulations of the full membrane structure, as shown in \S \ref{numerics}. We anticipate that our results could be used to generate reduced continuum models to describe diffusion problems in complex geometries such as examples explored in \cite{cherry2025boundary}, where the authors propose new numerical methods for these problems, exploiting Laplace transforms and boundary integral methods.

In this paper, our primary application of interest was the outer membrane of gram-negative bacteria. Using our steady effective conditions, \eqref{fullsteadcon}, we explored how the microstructure of this membrane affects its resulting permeability in \S \ref{permregimes}. With typical parameter values from a range of gram-negative bacteria, as given in Table \ref{tab:par}, we find that permeability is likely to be dominated by the membrane thickness, i.e. channel length, but hypothesise that reduced porin expression could lead to channel entrance effects becoming the limiting factor as porins become sparse. Both of these behaviours are captured in \eqref{steadperm}, from which we can obtain the appropriate form for the effective permeability in each limit, as discussed in \S \ref{permregimes}, and which can be used to calculate permeabilities specific to individual bacterial species. We also provide an explicit form for the effective permeability in the limit of $\mathcal{O}(1)$ aspect ratio channels in \eqref{aspectperm}, for example representing thinner membranes or wider porins, and calculate the appropriate effective conditions to use when the concentrations inside and/or outside the cell are time-varying in \labelcref{timeconcdiff,fluxdiff}.

It would be interesting to consider different membrane geometries and asymptotic limits in the future. For example, in \cite{benn_phase_2021} the authors found that porins form clustered patches in the outer membrane of \textit{E. coli}, suggesting a membrane geometry with periodic patches of porin-free and porin-dense regions is biologically relevant. This could also potentially be explored by adapting the techniques of \cite{crowdy_slip_2021}, where the author considered periodic windows of slots in a superhydrophobic grooved surface. More generally, the limit in which the width of our impermeable chunks goes to zero ($r^{*} - 2a^{*} \rightarrow 0$) could be of mathematical interest, as it may allow for fully analytical solutions of the problem, by extending the conformal mapping techniques used in \cite{crowdy_slip_2021,miyoshi_longitudinal_2022,miyoshi_fully_2024} to the domains here. 

In this paper, we considered membrane transport in 2D for simplicity. Through this we have identified key parameter groupings which determine the limiting factor for membrane transport. An analysis of 3D membrane transport would be an interesting extension to this work. We anticipate that while similar types of parameter groupings may determine the qualitative behaviour of membrane transport in 3D, the shape of the porin entrance and 3D geometry will play important roles. We therefore expect different functional forms of the membrane permeability in terms of membrane geometry; notably, the logarithmic singularity in \labelcref{steadperm,aspectperm} is likely a 2D phenomenon. More generally, one would not be able to rely on complex conformal maps to solve our inner regions in 3D.

Our physical parameter values, given in Table \ref{tab:par}, imply that cell radius is much larger than the membrane thickness, meaning curvature is small for spherical bacteria. Although we have neglected membrane curvature altogether in our model, we expect our derived coupling conditions, \labelcref{fullsteadcon,aspectperm,timeconcdiff,fluxdiff}, to hold for membranes with non-zero curvature, provided the curvature is not too large as to be observable in our boundary layer and inner regions. Similar leading-order independence of curvature is seen in \cite{gomez_asymptotic_2015}, when looking at asymptotic solutions of mean first passage time problems in non-spherical 3D domains, and in \cite{paquin2021modelling}, where the authors consider concentration differences between narrow windows on curved surfaces generated by localised diffusive fluxes. We note that, for a weakly curved membrane, it should be possible to systematically obtain higher-order corrections to our coupling conditions which explicitly depend on curvature, for example by using curvilinear coordinates in our boundary layer problems \ref{BLsec}, representing a small perturbation to flat space e.g. \cite{sbragaglia_note_2007}.

Furthermore, although we present our model in the context of a concentration, $c$, the problem set-up can be reframed in terms of probability density functions describing particle position, with only minor modifications to the subsequent analysis. In particular, one can consider \eqref{dimfull1} as a drift-free Fokker-Planck equation for the position of the centre of a finite-sized particle, and reframe a modified membrane geometry as the effective space in which the particle can move. As a result, we anticipate these results could be adapted in future work to describe transport of finite-sized particles, with some differences introduced from particle shape and interactions altering the effective geometry.

The effective coupling conditions that we have derived, \labelcref{fullsteadcon,aspectperm}, and \labelcref{fluxdiff,timeconcdiff}, for the steady and unsteady problems respectively, could be applied to the even larger scale of bacterial colony dynamics, if combined with upscaling methods such as \cite{dalwadi_upscaling_2018,dalwadi_systematic_2020}, to obtain a fully multiscale model that systematically accounts for membrane-level effects. In this paper we focused on membrane geometry parameters relevant to  bacterial membranes, however transport across membrane-type structures is important in many other contexts, including nutrient transport, waste removal, mass transfer through surface coatings in chemical engineering, water vapour loss through fabrics and carbon dioxide absorption on plant leaves via stomata \cite{stillwell_membrane_2016,wakeham_diffusion_1979,pappenheimer_passage_1953,baker2014gas}. The results in this work therefore have a much wider reach of applications, and may be used to understand diffusive transport through periodic narrow gaps in many contexts.

\section*{Acknowledgements and funding}
E.F.Y is supported by an EPSRC National Fellowship in Fluid Dynamics [EP/X027902/1], P.P. is supported by a UKRI Future
Leaders Fellowship [MR/V022385/1]. For the purpose of open access, the authors have applied a Creative Commons Attribution (CC BY) licence to any Author Accepted Manuscript version arising from this submission.

\section*{Data availability}
\noindent The computational tools are freely available at https://github.com/MollyB06/2d-effective-perm-paper.

\bibliographystyle{abbrv}
\bibliography{LT}
\appendix
\section{$\mathcal{O}(1)$ aspect ratio derivation} \label{o1details}
\begin{figure}
    \centering
    \includegraphics[width=0.8\linewidth]{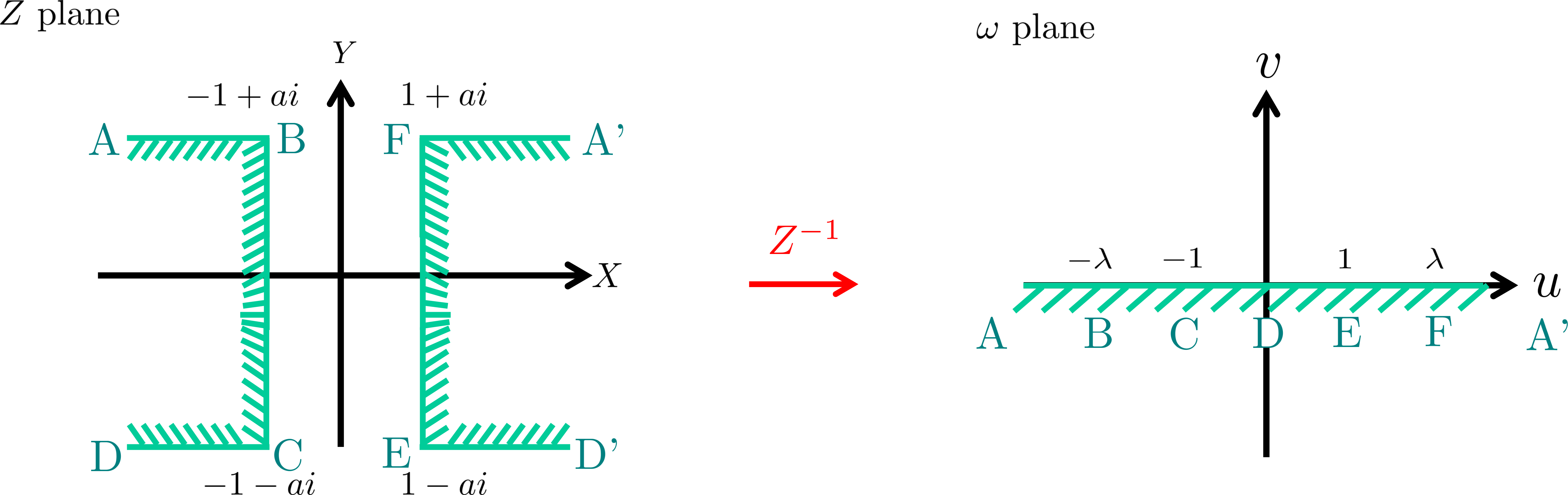}
    \caption{Schematic showing the result of the inverse of the conformal map \ref{aspectconfmap}, transforming our physical gap domain in $Z$ to the upper half-plane in $\omega$. The variable $\lambda$ describing the preimage of the point $1+ai$ under the map $Z$ is determined by the geometry and given implicitly in \eqref{lambdaimplicit}.}
    \label{fig:equalordermap}
\end{figure}

In this Appendix, we derive the $\mathcal{O}(1)$ aspect ratio effective permeability, \eqref{aspectperm}. As discussed in \S \ref{o1limit} the main change from the long thin channel limit is the coalescence of the three inner regions \mbox{III}a-c, into a single inner region. As in \S \ref{longthininner}, we start by scaling into this inner region using the transformations
\begin{equation}
    x = \delta \eps X, \hspace{1cm} y = \delta \eps Y, \label{o1cotrans}
\end{equation}
which converts \eqref{nondimgoverning} into the scaled system
\begin{subequations}
\label{o1innereq}
    \begin{align}
    \frac{\partial^2 \conc}{\partial X^2} + \frac{\partial^2 \conc}{\partial Y^2} &= 0, \quad \text{for } \{X \in \mathbb{R}, |Y| >a\} \cup \{|X|<1, |Y|<a\}, \label{o1eq1}\\
    \frac{\partial \conc}{\partial n} &= 0,\quad \text{for } \{|X|>1, |Y|=a\} \cup \{|X|=1,  |Y|<a\}.   
    \end{align}
\end{subequations}
We again only-require the leading-order problem and work directly with \eqref{o1innereq}, using a conformal map to simplify our domain. Setting $Z = X + iY$, we follow a similar procedure to \S \ref{inner1and3}, but for a different mapping. The map between the physical domain and the upper half-plane is shown in Figure \ref{fig:equalordermap}, which we can then straightforwardly map to the semi-infinite strip in which we solve our problem.
The Schwarz-Christoffel map from upper-half plane to the physical domain is
\begin{equation}
    Z = X + iY = f(\omega) = \tilde{P} + \tilde{d} \int_{a}^{\omega}\dfrac{(u^2-1)^{\frac{1}{2}}(u^2-\lambda^2)^{\frac{1}{2}}}{u^2} \mathrm{d}u.\label{schwarzint1}
\end{equation}
The constant $\lambda$ is defined as the preimage of the point $1+ai$ in the $Z$ domain and is determined implicitly. We can rewrite \eqref{schwarzint1} in a more interpretable form using the change of variables $t=1/u$, integrating by parts, and redefining the constant terms to obtain
\begin{equation}
    Z= P + d\left\{\omega \left(1-\dfrac{1}{\omega^2}\right)^{\frac{1}{2}} \left(1-\dfrac{\lambda^2}{\omega^2}\right)^{\frac{1}{2}} + \int_0^{\frac{1}{\omega}}\dfrac{(1-\lambda^2t^2)^{\frac{1}{2}}}{(1-t^2)^{\frac{1}{2}}}\mathrm{d}t + \lambda \int_{0}^{\frac{\lambda}{\omega}}\dfrac{(1-\frac{1}{\lambda^2}t^2)^{\frac{1}{2}}}{(1-t^2)^{\frac{1}{2}}}\mathrm{d}t \right\},\label{schwarzmap2}
\end{equation}
 where $P,d$ are the new constants. Our integral terms in \eqref{schwarzmap2} are now in the form of incomplete elliptic integrals of the second kind. We must take care when choosing our branch cuts across all of the square root terms in \eqref{schwarzmap2}. In particular, because $\omega$ lies in the upper half-plane, $1/\omega$ is in the lower half-plane. Hence, we choose our branch cuts to extend into the upper half-plane from the points $-1,-1/\lambda, 1/\lambda,1$ on the real axis, thus avoiding intersections with the physical domain. Using the desired geometry of our physical domain in $Z$ and the symmetry of the problem we set our unknown constants $P$ and $d$ to arrive at \eqref{aspectconfmap} as given in \S \ref{o1limit}, with $d$ defined in \eqref{o1const}. The dependence of $\lambda$ on $a$ is implicitly given through the relation \eqref{lambdaimplicit}.

As in \S \ref{inner1and3}, we further map into the semi-infinite strip to solve our problem and then transform back into the upper half-plane to write our solution implicitly in terms of $\omega$, obtaining
\begin{equation}
    \conc \sim \dfrac{2 b}{\pi}\log{|\omega|} + e. \label{aspectinnersol}
\end{equation}
Since \eqref{aspectconfmap} is not directly invertible, we use its limiting form as $|\omega| \rightarrow 0, \infty$ to match into our boundary layer regions by writing \eqref{aspectinnersol} in terms of $X$ and $Y$ . To match with Region \mbox{IV}, the boundary layer below the membrane, we take the limit $|\omega| \rightarrow 0$. In this limit \eqref{aspectconfmap} and \eqref{aspectinnersol} take the form
\begin{align}
\label{aspectclimbelow}
    Z \sim - \dfrac{d \lambda}{\omega}, \quad     c \sim \dfrac{2b}{\pi} \log{|d\lambda|} - \dfrac{2b}{\pi}\log{|Z|} + e \qquad \text{as } |\omega| \rightarrow 0.
\end{align}
Similarly to match into Region \mbox{II}, the boundary layer above the membrane, we take the limit as $|\omega| \rightarrow \infty$, obtaining the limiting forms of \eqref{aspectconfmap} and \eqref{aspectinnersol}:
\begin{align}
\label{aspectlimabove}
    Z \sim - d\omega, \quad     \conc \sim \dfrac{2b}{\pi} \log{|Z|} - \dfrac{2b}{\pi} \log{|d|} +e
    \qquad \text{as } |\omega| \rightarrow \infty.
\end{align}
Finally, we also match \labelcref{aspectclimbelow,aspectlimabove} to our boundary layer solutions, \labelcref{steadsolbound1,steadsolbound2}, in the limits $Y_{1,2} \rightarrow 0_{-,+}$ to obtain:
    \begin{align}
        \kappa_1 = -\kappa_2,\quad
        A_1 = A_2 + \dfrac{\kappa_2}{\pi}\log{\eps^2 \pi^2 \lambda d^2}.
    \end{align}
Using \eqref{outertoboundrelations}, the previous matching between the boundary layer regions and outer regions, to give $\kappa_{1,2}$ and $A_{1,2}$ in terms of the outer concentrations and fluxes, we obtain our $\mathcal{O}(1)$ aspect ratio coupling conditions with permeability \eqref{aspectperm}.

\section{Slow convergence in the early-time behaviour} \label{EMappend}
As noted in \S 4(b), we can circumvent slow convergence in the infinite sums in (4.12) and (4.13) as $t\rightarrow0^{+}$ by using an asymptotic Euler-Maclaurin approximation to convert the slowly converging sums into integrals.
For simplicity, we assume vanishing initial conditions (i.e $g_n(0)=0$) although our results are straightforward to extend to other initial conditions. Essentially, our goal is to approximate the sums:
\begin{subequations}
    \begin{align}
      &\sum_{n \text{ odd}}  \exp\left(- \frac{\lambda_n^2 \ndtime}{4}\right)\int_{0}^{\ndtime} \exp\left(\frac{\lambda_n^2 \tau}{4}\right)\frac{\partial}{\partial \tau}\left(h(\tau) \right)\; \mathrm{d} \tau \label{odd sum}, \\
      & \sum_{n \text{ even}} \exp\left(- \frac{\lambda_n^2 \ndtime}{4}\right)\int_{0}^{\ndtime}\exp\left(\frac{\lambda_n^2 \tau}{4}\right)\frac{\partial}{\partial \tau} \left(f(\tau) \right)\; \mathrm{d} \tau \label{even sum},
    \end{align}
\end{subequations}
where
\begin{subequations}
    \begin{align}
        &h(\tau)= \frac{\concabove + \concbelow}{\chanL} - \frac{\delta }{\chanL\pi}\left(\log{\frac{1}{8 \eps}}+ 1 \right) \left( \frac{\partial \concabove}{\partial y}-  \frac{\partial \concbelow}{\partial y}\right), \label{htau}\\
        & f(\tau) = \frac{\concabove - \concbelow}{\chanL} - \frac{\delta }{\chanL\pi}\left(\log{\frac{1}{8 \eps}}+ 1 \right)\left(\frac{\partial \concabove}{\partial y}+  \frac{\partial \concbelow}{\partial y} \right) \label{ftau}.
    \end{align}
\end{subequations}
We outline the method for the odd sum \eqref{odd sum} below, as the result for the even sum \eqref{even sum} follows equivalently. 

Since the integration domain is small for early time, we first expand $h(\tau)$ around $\tau = t$, allowing us to rewrite \eqref{odd sum} as
\begin{equation}
    \sum_{n \text{ odd}}  \exp\left(- \frac{\lambda_n^2 \ndtime}{4}\right)\int_{0}^{\ndtime} \exp\left(\frac{\lambda_n^2 \tau}{4}\right)\frac{\partial}{\partial \tau}\left(h(\tau) \right)\; \mathrm{d} \tau 
 \sim \sum_{\text{n odd}} \frac{4}{\lambda_n^2} h'(t)\left(1 -  \exp\left(-\frac{\lambda_n^2 t}{4}\right) \right). \label{reduced odd sum}
\end{equation}
Using $\lambda_n = n \pi / \chanL$ and $\sum_{\text{n odd}} n^{-2} = \pi^2/{8}$, we can write \eqref{reduced odd sum} as
\begin{equation}
      \sum_{\text{n odd}} \frac{4}{\lambda_n^2} h'(t)\left(1 -  \exp\left(-\frac{\lambda_n^2 t}{4}\right) \right) = \frac{h'(t) \chanL^2}{2}-\frac{4 h'(t) \chanL^2}{\pi^2} \sum_{\text{n odd}}\frac{\exp\left(-\left(\frac{\pi^2}{4\chanL^2} \right)n^2 t\right)}{n^2}. \label{new odd sum}
\end{equation}
To evaluate the sum in \eqref{new odd sum}, we use an asymptotic Euler-Maclaurin approximation. In order for the remainder terms in this approximation to decay asymptotically, we split the summation as follows
\begin{equation}
   \sum_{\text{n odd}}\frac{\exp\left(-\left(\frac{\pi^2}{4\chanL^2} \right)n^2 t\right)}{n^2} = \sum_{m =1 }^{p-1}\frac{\exp\left(-\left(\frac{\pi^2}{4\chanL^2} \right)(2m-1)^2 t\right)}{(2m-1)^2} + \sum_{m = p}^{\infty}\frac{\exp\left(-\left(\frac{\pi^2}{4\chanL^2} \right)(2m-1)^2 t\right)}{(2m-1)^2}, \label{split sum}
\end{equation}
where we choose $p$ such that $1 \ll p \ll 1/\sqrt{t}$. Taylor expanding the first term and using an Euler-Maclaurin expansion on the second, \eqref{split sum} becomes
\begin{equation}
   \sum_{\text{n odd}}\frac{\exp \left(-\left(\frac{\pi^2}{4\chanL^2} \right)n^2 t\right)}{n^2} \sim \sum_{m=1}^{p-1} \frac{1}{(2m-1)^2} + \frac{\pi\sqrt{t}}{4 \chanL} {\int}_{\frac{\pi \sqrt{t}}{2 \chanL}(2p-1)}^{\infty} \dfrac{\exp\left(-s^2\right)}{s^2} \mathrm{d}s + \frac{\exp\left(-(2p-1)^2 t \frac{\pi^2}{4 \chanL^2}\right)}{2(2p-1)^2}. \label{approxsumodd} 
\end{equation}
Taking $p \rightarrow \infty$, with $\sqrt{t} p \rightarrow 0$ in \eqref{approxsumodd} we obtain
\begin{equation}
\sum_{\text{n odd}}\frac{\exp\left(-\left(\frac{\pi^2}{4\chanL^2} \right)n^2 t\right)}{n^2} \sim \frac{\pi^2}{8} - \frac{\sqrt{\pi^3t}}{4 \chanL} \label{simpsumodd}. 
\end{equation}
Then combining \eqref{simpsumodd} and \eqref{new odd sum}, we deduce
\begin{subequations}
\label{resultsums}
\begin{align}
        \sum_{\text{n odd}} \frac{4}{\lambda_n^2} h'(t)\left(1 -  \exp\left(-\frac{\lambda_n^2 t}{4}\right) \right) \sim h'(t) \chanL \sqrt{\dfrac{t}{\pi}}. \label{resultoddsum}
\end{align}
In a similar manner, we can approximate \eqref{even sum} at early times as
\begin{align}
    \sum_{n \text{ even}} \exp\left(- \frac{\lambda_n^2 \ndtime}{4}\right)\int_{0}^{\ndtime}\exp\left(\frac{\lambda_n^2 \tau}{4}\right)\frac{\partial}{\partial \tau} \left(f(\tau) \right)\; \mathrm{d} \tau \sim f'(t)\chanL \sqrt{\dfrac{t}{\pi}}. \label{resultevensum}
\end{align}
\end{subequations}
Using \labelcref{resultsums}, at early times we can rewrite our coupling conditions (4.12) and (4.13) from \S 4(b), as (4.14), thereby circumventing the slow convergence.

\section{Accuracy of coupling conditions} \label{app:error}
\begin{figure}
    \begin{overpic}[scale=0.55,percent ,grid=false]{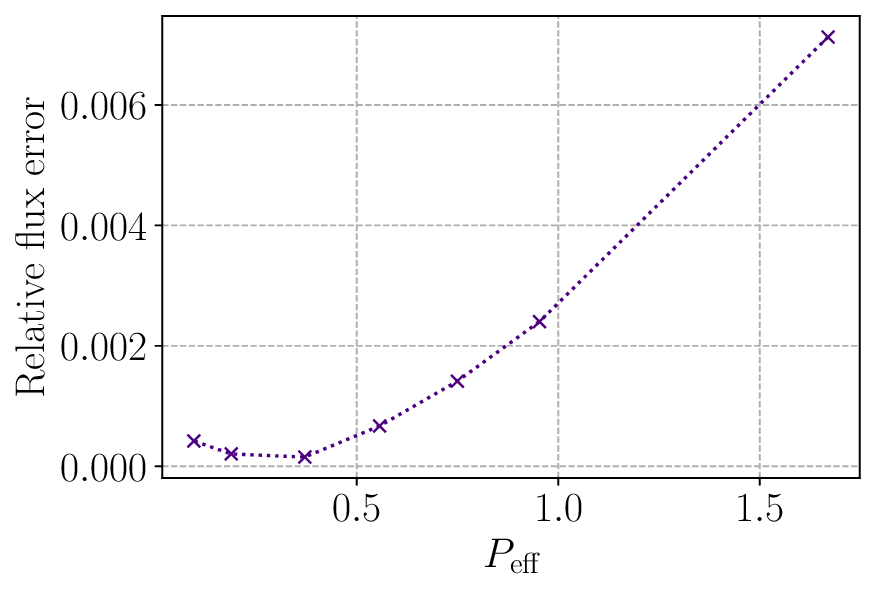}
    \put(5,70){(a)}
    \put(102,0){\includegraphics[scale=0.55]{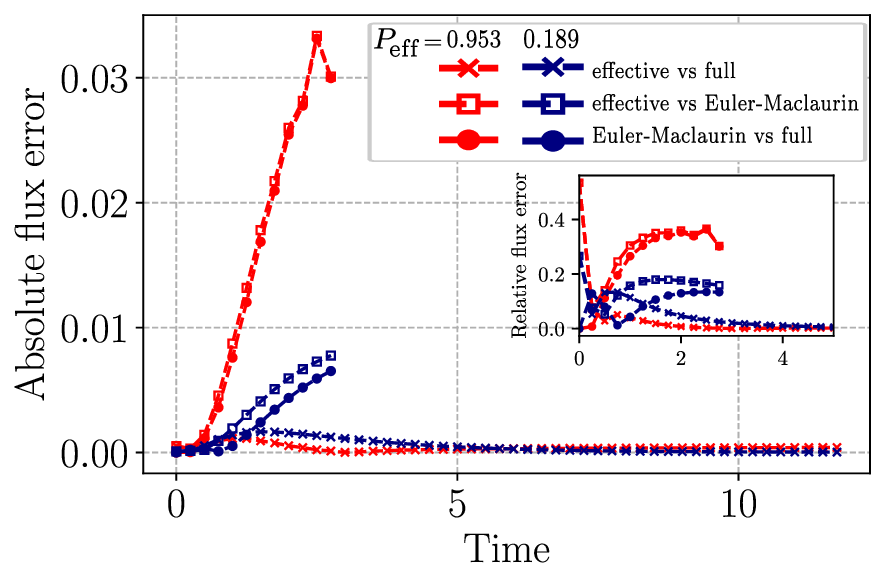}}
    \put(102,70){(b)}
    \end{overpic}
    \caption{Error in average flux when using the coupling conditions compared to the flux obtained from simulating the full membrane geometry. In both cases flux is calculated over a horizontal line away from the membrane. (a) Relative flux error vs effective permeability, (3.35b), in the steady problem. We plot the error between the analytical flux obtained from (5.1) and a simulation of the full membrane geometry. Permeability is varied by fixing $\delta = 0.25$ and varying $\eps$, with similar results obtained by fixing $\eps$ and varying $\delta$.
    (b) Absolute flux error vs time in the unsteady problem for two different membrane geometries with steady effective permeabilities of 0.189 (blue) and 0.953 (red), inset shows relative flux error. We plot the error between a simulation of the effective interface condition with the infinite sums truncated at 30 terms, (4.12) and (4.13), and the full membrane geometry (crosses), the error between simulations of the effective interface condition and the Euler-Maclaurin approximation for the interface condition, (4.14), up to $t=2.75$ (circles) and the error between simulations of the full membrane geometry and the Euler-Maclaurin approximation of the effective interface condition up to $t=2.75$ (squares).\\}
    \label{relfluxerrors}
    \end{figure}

\begin{figure}[t]
\begin{overpic}[abs,unit=1mm,scale=0.3,grid=false]{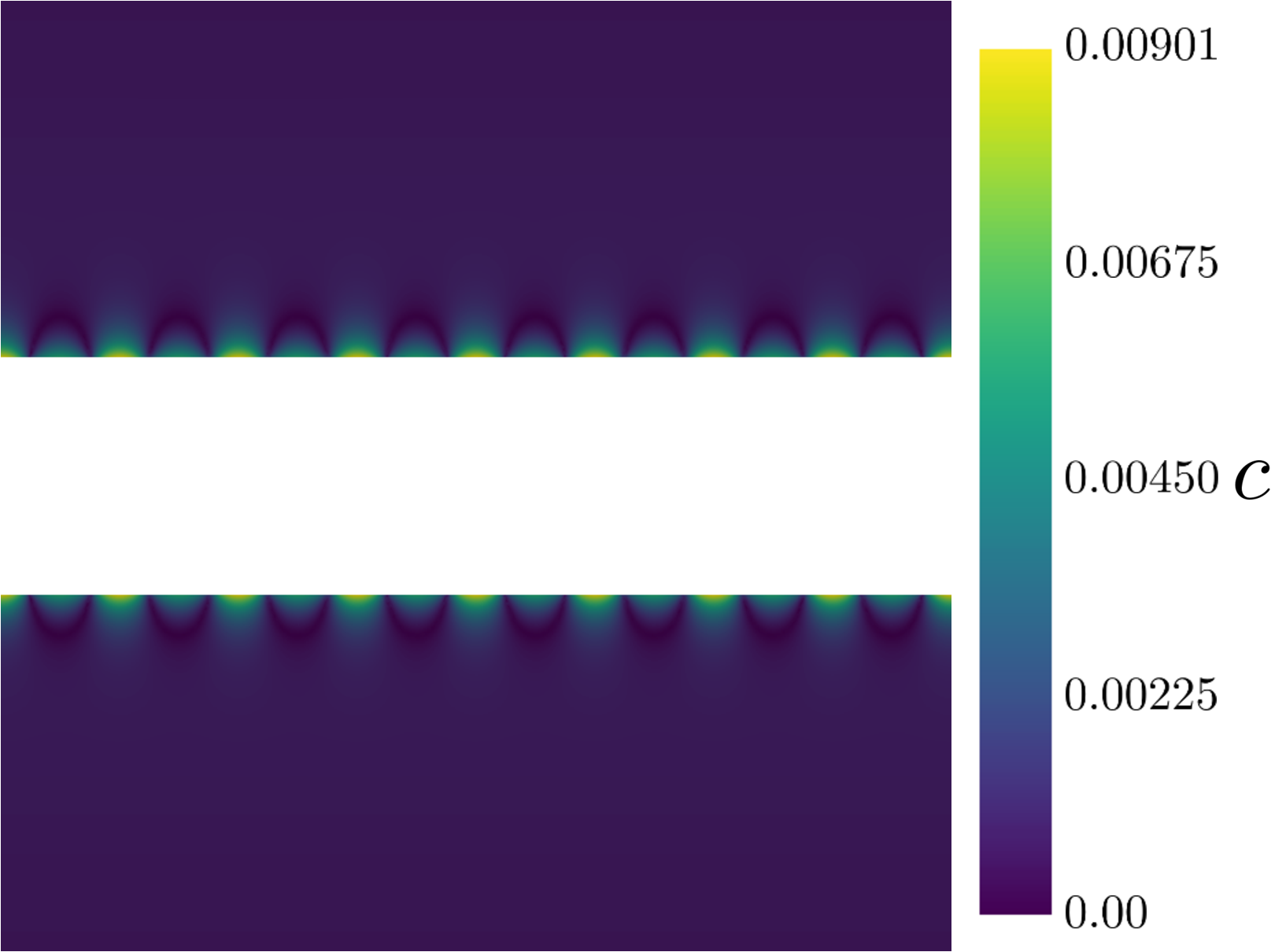}
\put(0,65){(a)}
\put(85,0){\includegraphics[scale=0.3]{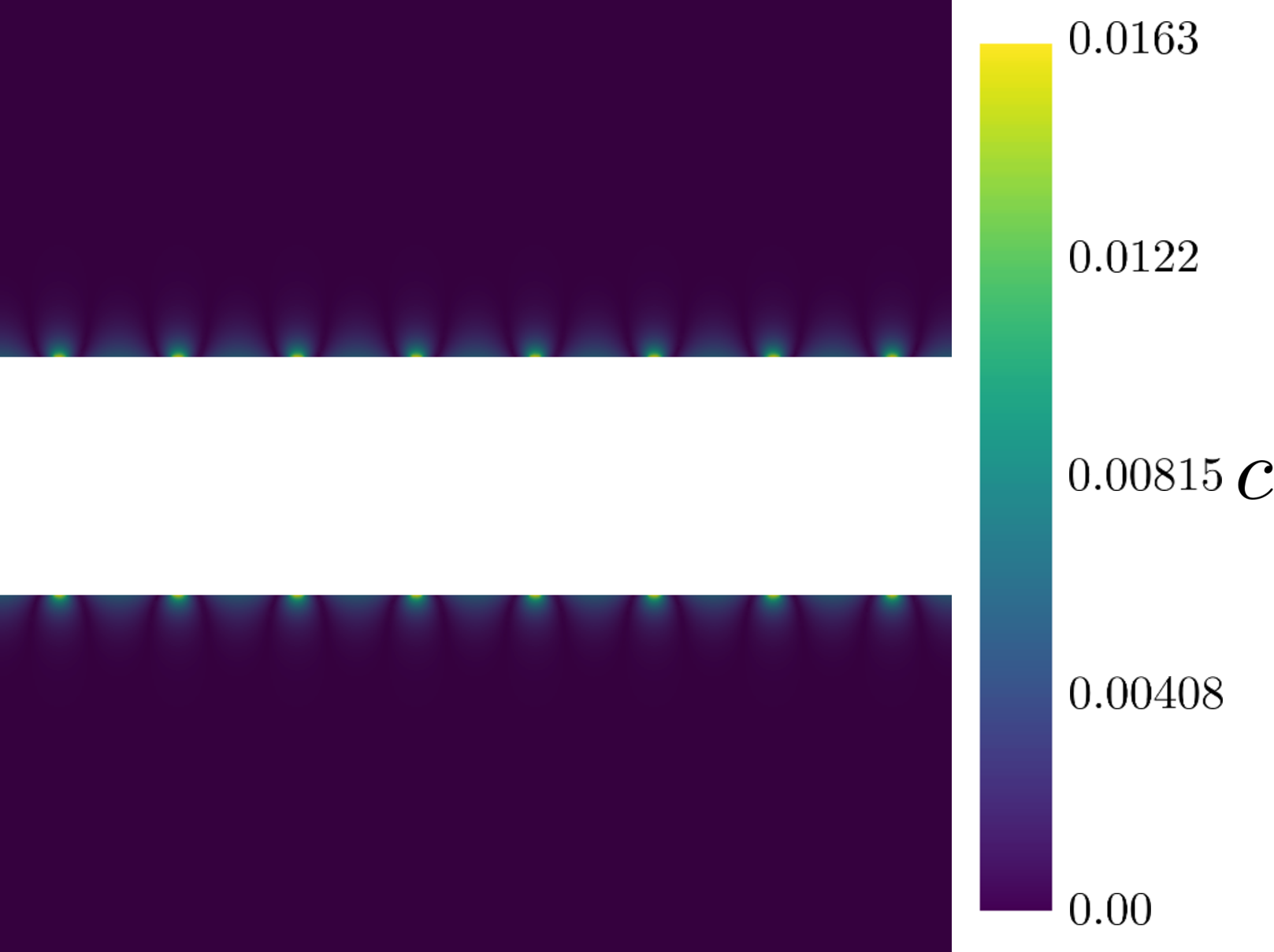}}
\put(83,65){(b)}
\end{overpic}
    \caption{Absolute difference between the full and effective problem concentration profiles for a membrane with effective permeabilities (a) $P_{\text{eff}} = 0.953$, and (b) $P_{\text{eff}} =0.188$. The absolute difference is only non-negligible close to the membrane. In order to visualise these errors, we have zoomed into a 2x2 region in the centre of our original 5x5 domains (noting that the full 5x5 domains are presented in Figures 7 and 8).\\}
    \label{absdiff}
\end{figure}

In this section we demonstrate the accuracy of the coupling conditions for both the steady and unsteady problem by reporting the error in the average fluxes over a horizontal line away from the membrane in Figure \ref{relfluxerrors} as well as visualising the difference in concentration profiles shown in Figures 7 and 8 in \ref{absdiff}. This error corresponds to the difference between the flux curves in Figure 9. Panel (a) shows relative flux error vs $P_{\text{eff}}$ in the steady problem and panel (b) shows flux error vs time in the unsteady problem, where in the main figure we show absolute  flux error vs time with an inset showing relative flux error vs time. In Figure \ref{relfluxerrors} (a) we look at the relative error in average flux between the analytical solution of our effective interface problem and a simulation of the full membrane geometry. The increase in relative error for small permeabilities is due to the very narrow channels not being resolved properly with the fixed mesh size used. Accuracy at these smaller permeabilities could be improved by the use of a finer mesh. In Figure \ref{relfluxerrors} (b) we show the absolute errors between the full geometry and full effective condition, the full geometry and the Euler-Maclaurin approximation for the effective interface condition and the absolute error between the full effective interface condition and Euler-Maclaurin approximation for this condition. As in Figure 9 we terminate the Euler-Maclaurin simulations at $t=2.75$ as they start to deviate more significantly from the small-time effective problem they are approximating after this point. We include an inset showing relative flux error for each of the cases mentioned above, noting the larger relative errors at early times are a result of dividing by very small fluxes.
\begin{figure}
    \begin{overpic}[scale=0.55,percent ,grid=false]{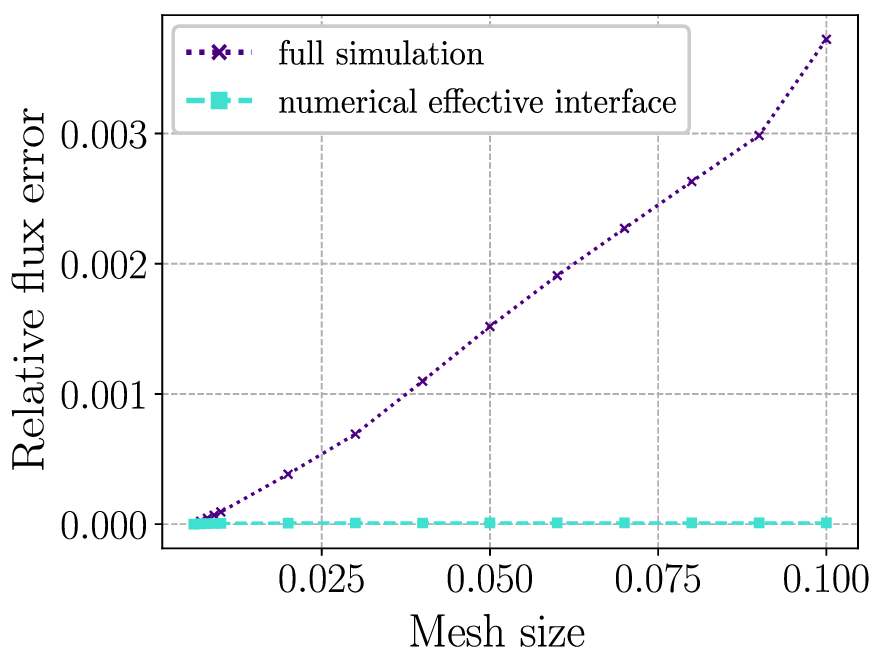}
    \put(5,74){(a)}
    \put(102,0){\includegraphics[scale=0.58]{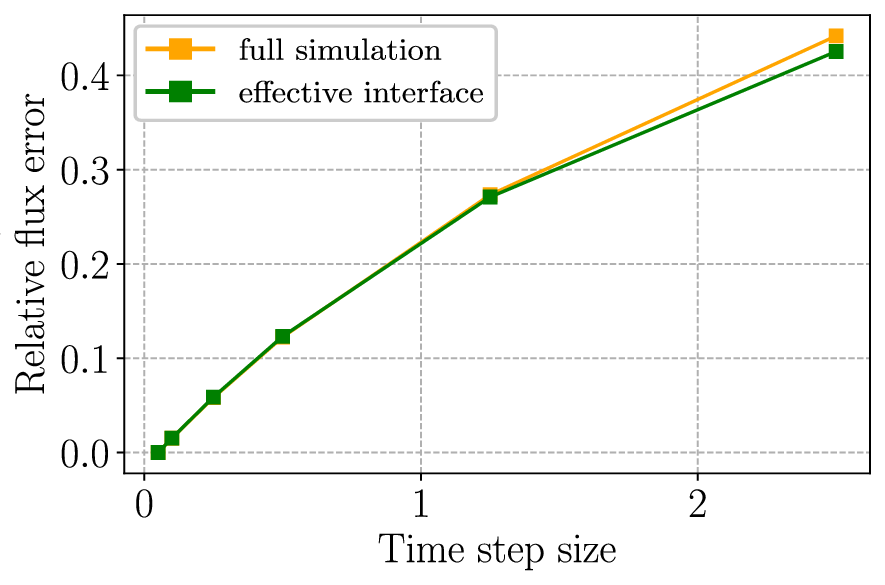}}
    \put(102,74){(b)}
    \end{overpic}
    \caption{(a) Relative error in average flux over a horizontal line away from the membrane vs mesh size for simulations of the effective interface (blue squares) and the full geometry (purple crosses). Relative error is calculated using \eqref{equation:relative_er}.
    (b) Relative error in average flux over a horizontal line away from the membrane vs time step size. Relative flux error is calculated using \eqref{equation:er-time}.}
    \label{numconv}
\end{figure}

\section{Numerical convergence}\label{appendix:numerical}
In this section we confirm the convergence of our finite element scheme for the full problem and effective interface condition, using the error obtained relative to simulations using the smallest mesh size or time step size, respectively. We confirm the convergence of our steady simulation in the long thin channel limit as we decrease the mesh size, $h$, in Figure \ref{numconv}(a). Mesh size is defined as the target size of the elements in the mesh, and can be uniform or adaptive \cite{geuzaine2009gmsh}. We calculate relative error as \begin{align}
\text{Er}=\frac{|\text{flux}_{\text{mesh size = h}} - \text{flux}_{\text{mesh size = 0.007}}|}{\text{flux}_{\text{mesh size = 0.007}}}\label{equation:relative_er},
\end{align}for a range of mesh sizes, $h \in[0.007,0.1]$. We find that the relative error in average flux over a horizontal line away from the membrane is below. For simulations of our effective condition we use a uniform mesh with mesh size $h=0.008$. To help resolve the narrow channels in simulations of our full membrane geometry we use a spatially adaptive mesh with smallest mesh size $h=0.008/5$ and largest mesh size $5\times 0.008$.

We confirm the convergence of our unsteady simulation as the time-step size $\Delta t$ decreases in Figure \ref{numconv}(b). We measure the relative error in the average flux at $t=2.5$
for different time-step sizes using \begin{align}
   \text{Er}=\frac{|\text{flux}_{\Delta t=s} - \text{flux}_{\Delta t = 0.05}|}{{\text{flux}_{\Delta t=0.05}}}\label{equation:er-time},
\end{align}
with $s\in[0.05,2.5]$.
We obtain a relative error below 10\% for $\Delta t$ just under $0.5$, as a result we select the time-step $s=0.25$, to balance numerical convergence with simulation time and computational memory. This is in part due to our storing the concentration value at each time-step in order to compute the memory integrals in (4.12) and (4.13), using a numerical quadrature. We anticipate this could be improved in future work using a time-marching scheme, similar to that presented in \cite{pelz2025synchronized}.

\end{document}